\documentclass[a4paper,fleqn]{mnras}

\usepackage[T1]{fontenc}
\usepackage{ae,aecompl}

\usepackage{graphicx}	
\usepackage{amsmath}	
\usepackage{amssymb}	

\title[An extensive radial velocity survey toward NGC~6253]{An extensive radial velocity survey toward NGC~6253}

\author[M. Montalto et al.]{
M. Montalto,$^{1}$\thanks{E-mail: marco.montalto@astro.up.pt}
C. H. F. Melo,$^{2}$
N. C. Santos,$^{1,3}$
D. Queloz,$^{4}$
G. Piotto,$^{5}$
S. Desidera,$^{6}$
\newauthor
L. R. Bedin,$^{6}$
Y. Momany,$^{6}$
I. Saviane$^{2}$
\\
$^{1}$Instituto de Astrof\'isica e Ci\^encias do Espa\c{c}o, Universidade do Porto, CAUP, Rua das Estrelas, PT4150-762 Porto, Portugal\\
$^{2}$European Southern Observatory, Alonso de Cordova 3107, Vitacura Casilla 19001, Santiago 19, Chile\\
$^{3}$Departamento de Fisica e Astronomia, Faculdade de Ci\^encias, Universidade do Porto, Portugal\\
$^{4}$Observatoire de Gen\'eve, 51 Ch. des Mailettes, 1290 Sauverny, Switzerland\\
$^{5}$Dipartimento di Fisica e Astronomia Galileo Galilei, Universit\'a di Padova, Vicolo dellOsservatorio 3, Padova IT-35122\\
$^{6}$INAF - Osservatorio Astronomico di Padova, Vicolo dellOsservatorio 5, Padova, IT-35122
}

\date{Accepted XXX. Received YYY; in original form ZZZ}

\pubyear{2015}

\begin{document}
\label{firstpage}
\pagerange{\pageref{firstpage}--\pageref{lastpage}}
\maketitle

\begin{abstract}
The old  and metal rich  open cluster NGC  6253 was observed  with the
FLAMES multi-object  spectrograph during an  extensive radial velocity
campaign monitoring 317  stars with a median of  15 epochs per object.
All the  targeted stars are located  along the upper  main sequence of
the  cluster  between 14.8  $<$  V $<$  16.5.   Fifty nine stars  are
confirmed cluster members both by radial velocities and proper motions
and  do not  show evidence  of variability.   We detected  45 variable
stars among  which 25 belong to NGC  6253.  We were able  to derive an
orbital  solution  for 4  cluster  members  (and  for 2  field  stars)
yielding  minimum masses  in between  $\sim$90 M$\rm_J$  and $\sim$460
M$\rm_J$ and periods between 3 and 220 days.  Simulations demonstrated
that this  survey was sensitive to  objects down to 30  M$\rm_J$ at 10
days orbital periods with a detection efficiency equal to 50\%. On the
basis of  these results  we concluded that  the observed  frequency of
binaries down to the hydrogen burning  limit and up to 20 days orbital
period is  around (1.5$\pm$1.3)\% in  NGC 6253.  The  overall observed
frequency  of   binaries  around  the  sample  of   cluster  stars  is
(13$\pm$3)\%.  The  median radial  velocity precision achieved  by the
GIRAFFE   spectrograph   in    this   magnitude   range   was   around
$\sim$240m$\rm\,s^{-1}$ ($\sim$180 m$\rm\,s^{-1}$ for UVES).  Based on
a limited follow-up  analysis of 7 stars in our  sample with the HARPS
spectrograph we determined  that a precision of 35  m $\rm s^{-1}$ can
be  reached  in this  magnitude  range,  offering  the possibility  to
further extend  the variability  analysis into the  substellar domain.
Prospects  are  even more  favourable  once  considering the  upcoming
ESPRESSO spectrograph at VLT.
\end{abstract}

\begin{keywords}
open clusters: general -- open clusters: individual NGC~6253
\end{keywords}



\section{Introduction}
\label{sec:introduction}

NGC            6253            ($\alpha_{2000}=16^{h}\,59^{m}\,05^{s},
\delta_{2000}=-52^{\circ}\,42\arcmin\,30\arcsec,
l=335\fdg5,b=-6\fdg3$) is  an old and metal-rich  open cluster located
inside the  solar ring  and projected toward  a rich  Galactic stellar
field in the direction of  the Galactic center.  The metal-rich nature
of this cluster and its  importance in the context of Galactic studies
were early  recognized in the  seminal work of Bragaglia  et al.~(1997)
and later on discussed by several other authors in the past (Piatti et
al.  1998; Sagar,  Munari, \& de Boer 2001;  Twarog, Anthony-Twarog \&
De  Lee  2003; Anthony-Twarog,  Twarog,  \&  Mayer  2007; Carretta  et
al. 2000; Carretta,  Bragaglia \& Gratton 2007; Sestito  et al.  2007;
Anthony-Twarog  et  al.   2010).   Novel chemical  studies  have  been
recently presented in Mikolaitis  et al.~(2012) where C,N,O abundances
and carbon  isotope ratios for four  red clump stars  of NGC~6253 were
analyzed and in Cummings et al.~(2012), focused on lithium abundance.
  
In the  context of extrasolar  planet searches NGC~6253 plays  a major
role.  Given its high  metallicity and  the well  known fact  that the
frequency  of jupiter  planets found  around field  stars is  known to
strongly  correlate with metallicity  (e.g.  Gonzalez~1997;  Santos et
al.~2001; Santos et al.~2004; Fischer \& Valenti 2005; Mortier et al. 2012) we may expect a
large  population of  these objects  to  be present  in this  cluster,
unless  other  evolutionary  and  environmental  factors  alter  their
formation histories.

Our  previous  efforts  to   detect  this  population  and  to  better
characterize cluster properties were  presented in a sequence of works
starting with Montalto et al.~(2009, hereafter M09) where we delivered
a  photometric  and   astrometric  catalog  including  proper  motions
membership  probabilities.   De   Marchi  et  al.~(2010)  studied  the
photometric  variability   properties  of  cluster   members  and  the
surrounding field  producing a catalog  of 595 variables, 35  of which
were proposed as cluster  members. In Montalto et al.~(2011, hereafter
M11) we  presented the results  of a radial velocity  survey conducted
with the UVES/GIRAFFE spectrographs at VLT, mostly concentrated on the
evolved  portion of  the color-magnitude  diagram.  In  that  work, we
presented a follow-up analysis  of three planetary candidates we found
in our  photometric studies and  highlight the discovery of  the first
cluster double lined eclipsing  binary system, located in the turn-off
region (star 45368 in our catalog) which membership was confirmed both
by proper motions  and by radial velocities.  We  analyzed a sample of
139  stars in  the  cluster  region concluding  that  35 where  likely
cluster members and 12 likely close binary systems.

In Montalto  et al~(2012), we performed equivalent  width analysis and
derived Fe,  Si, Ca,  Ti, Cr  and Ni abundances  as well  as abundance
ratios  of  a main-sequence  star,  two  red-clump  stars and  a  blue
straggler cluster members.  For  our main-sequence star, we obtained a
metallicity of [Fe/H]=+0.26$\pm$0.11 (rms), whereas for the two giants
we found that our  metallicities were on average [Fe/H]=+0.19$\pm$0.13
(rms), lower than what was  determined in previous studies (Sestito et
al.~2007,   [Fe/H]=+0.36$\pm$0.07   rms,   Carretta  et   al.    2007,
[Fe/H]=+0.46$\pm$0.03 rms).

NGC-6253 fundamental  parameters are not  yet firmly set.  It  is well
known that  the different calibrated photometries presented  so far do
not agree well, and that significant differences exist among them. The
cluster  age  was  set  to  3.5  Gyr  in  M09  and  the  reddening  to
$E(B-V)$=0.15. These  estimates were respectively the  upper bound age
and the lower bound reddening  derived from isochrone fit so far, once
compared with  the other results in the  literature.  However recently
Rozyczka et al.~(2014) presented  a detailed analysis of the eclipsing
binary 45368 (renamed in their work V15).  Their results indicate that
the age of  NGC~6253 should be comprised between 3.80  - 4.25 Gyr from
the mass-radius diagram and should be  even older (3.9 - 4.6 Gyr) from
color-magnitude diagram  (CMD) fitting.  They also  report a reddening
equal to $E(B-V)$=0.113  mag, and therefore lower than  our own value.
The same authors  presented in Kaluzny et al.~(2014)  the results of a
novel photometric  campaign in the cluster region,  conducted with the
1.0m Swope Telescope in Las Campanas, focusing primarily on the bright
portion of the CMD, in a domain only partially covered by our previous
photometric surveys.   Among a sample  of 25 additional  variables the
authors  detected  three  novel  eclipsing  binaries  members  of  the
cluster.  The  analysis of  these objects will  likely  permit to
reach more stringent constraints  on cluster properties and to further
complement  the  results  obtained   for  45368.   It  has  been  also
demonstrated that theoretical models  fail to accurately reproduce the
observed    CMD,     particularly    for    evolved     stars    (M09,
Anthony-Twarog~2010, Rozyczka et al.~2014).
 
In  this work,  we present  the  results of  another extensive  radial
velocity  campaign that  was performed  on  this cluster  in the  past
years.  This survey  superseeds by  far  both in  number of  monitored
objects and in number of epochs the analysis presented in M11 and also
the similar  one discussed  in Anthony-Twarog~(2010). It  represents a
more   ambitious  effort  to   better  characterize   radial  velocity
variability among  cluster members, and  an important step  toward the
detection of low mass objects.  
The preparation  of the observations presented in  this work preceeded
in time the photometric and astrometric analysis discussed in
M11. Sample selection was based  on the work of Twarog, Anthony-Twarog
\& De Lee (2003). The magnitude range of the sample stars is comprised
in  between 14.8$<$  V $<$16.5.   All  the targeted stars lie on the upper
main sequence of NGC~6253.

In  Section~\ref{sec:observations},  we describe  the observations  we
acquired.      In  Section~\ref{sec:reduction},    we     discuss data
reduction. In Section~\ref{sec:search}, we present the methods used to
identify   variables and  in Section~\ref{sec:classification}      our
classification criteria.  In Section~\ref{sec:spectroscopic_binaries},
we describe  the analysis of a sample   of spectroscopic binaries.  In
Sec.~\ref{sec:comparison}, we compare   our  results  with  the  ones
obtained during  previous surveys.  In  Section~\ref{sec:simulations},
we  calculate  the     survey   detection  efficiency.    Finally   in
Section~\ref{sec:conclusions}, we summarize and conclude.


\section{Observations}
\label{sec:observations}

The observations described in this work have been acquired with FLAMES
(Pasquini  et  al.   2002)  and   HARPS  (Mayor,  M.   et  al.   2003)
spectrographs.  FLAMES  is  the  multi-object, intermediate  and  high
resolution  spectrograph  of the  VLT  installed  at  the UT2  (Kueyen
telescope), in Paranal, Chile.  FLAMES  is a complex system that feeds
two different  spectrographs, UVES  and GIRAFFE.  While  UVES provides
the  maximum resolution  (R =  47  000), but  can access  up to  eight
targets at  a time, GIRAFFE has an  intermediate resolution permitting
to target up  to 132 objects at the time  or to perform integral
field  spectroscopy.   FLAMES  data  were obtained  in  two  different
observing  seasons  between  April-  July, 2004  and  March-July  2005
respectively.  For UVES we used  the standard setup centered at 580nm.
For  GIRAFFE we used  different high  resolution settings:  HR8, HR9B,
HR11, HR12,  HR13, HR14A, HR14B overall covering  the wavelength range
between 491.7nm and 670.1 nm  with a resolution in between R=17740 and
R=28800.  In total 79 epochs were acquired obtaining 6558 spectra (553
with UVES and 5955 with GIRAFFE) corresponding to 317 stars.  In order
to allow  high accuracy in  the radial velocity measurements  for both
spectrographs  simultaneous  Th-Ar  lamps observations  were  acquired
together with the  scientific targets.  However, for 21  out of the 79
GIRAFFE plates  no simultaneous  calibrations were acquired.   For the
rest of the  plates 5 fibers were allocated  to the calibration lamps,
while  for  UVES 1  fiber  was  always  allocated to  the  simutaneous
calibration  lamp.   For   both  spectrographs  no  fibers  were
  allocated to the sky.

HARPS is the High Accuracy  Radial velocity Planet Searcher at the ESO
La  Silla 3.6m  telescope. We  used the  High Efficiency  mode (EGGS).
HARPS  data   have  been  obtained  between  28-30   May,  2011.   The
observations were  much more limited  with respect to the  FLAMES runs
and in  general obtained under  non optimal conditions. We  observed a
total of 7 stars and a maximum of 5 epochs per object, resulting in 20
measurements.  
These targets were  selected among the  sample of
stars earlier observed with UVES/GIRAFFE with the purpose to follow-up
potential planetary candidates. The early analysis of the UVES/GIRAFFE
data relied  upon  strong
assumptions on the precision of the measurements, which was claimed at
the level of 40-50 m/s.  Furthermore, such  analysis neglected
the  presence  of potential  and  very  poorly  understood sources  of
systematics  that   could  have   compromised  the  accuracy   of  the
measurements. In  this work, we  will study in detail  this problem,
presenting for the first time a complete and troughout analysis of the 
entire dataset.

\noindent
Table~\ref{tab:observations} offers  an overview of  all the observing
runs.

\section{Data analysis}
\label{sec:reduction}

UVES  data were  reduced using  the  REFLEX (Freudling  et al.   2013)
UVES-FIBER workflow.   GIRAFFE data  were reduced using  GASGANO.  For
HARPS data we retrieved from the ESO archive the Science Data Products
via     the     Phase      3     spectral     query     
form\footnote{http://archive.eso.org/wdb/wdb/adp/\\phase3\_spectral/form?phase3\_collection=HARPS}.

\begin{figure*}
\centering
\includegraphics[width=16cm]{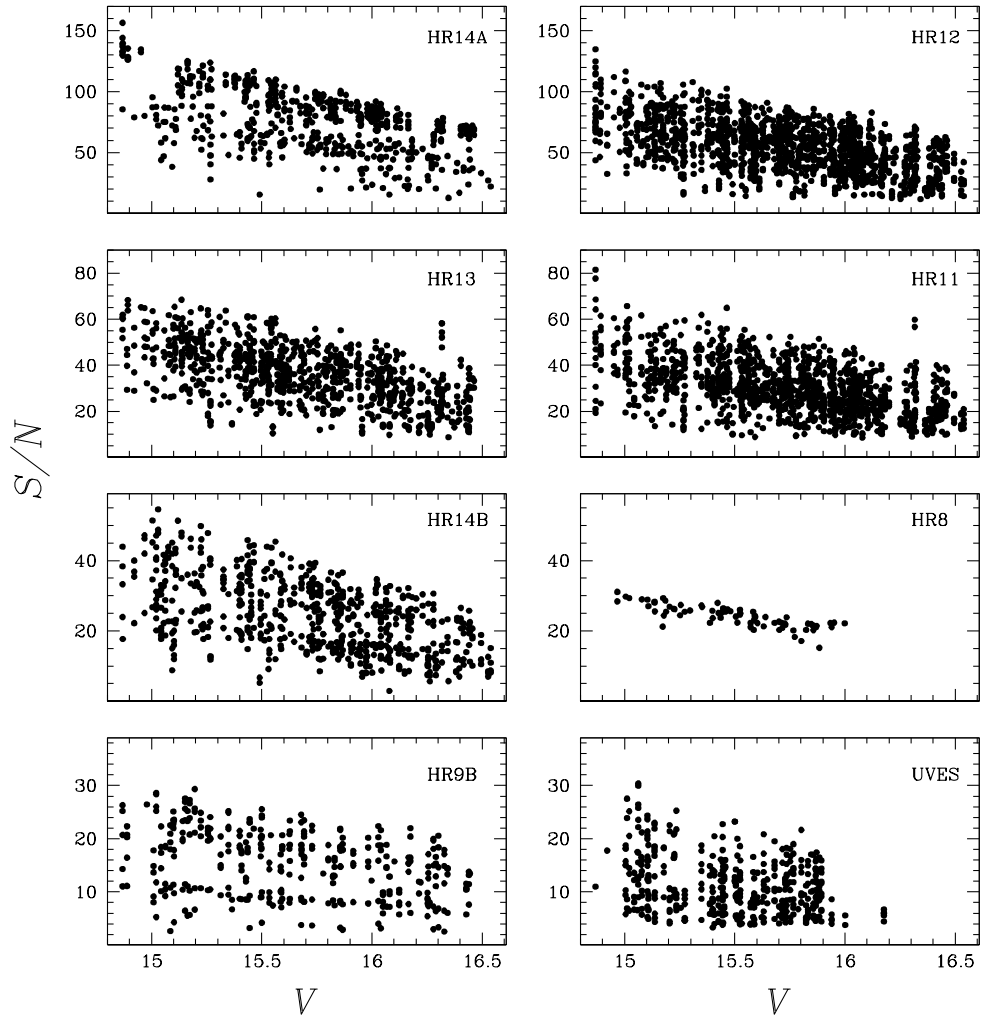}
\caption{
Observed S/N for  the spectra acquired with each  GIRAFFE setting used
in this work and with the UVES spectrograph (bottom right panel).
\label{fig:sn}
}
\end{figure*}

\noindent
After  the  data reduction  step  we  cross-correlate  all the  FLAMES
spectra  with a  reference  spectrum.  Because  all  of the  monitored
objects were located on the upper main sequence of the CMD (see below)
we adopted a solar spectrum as  a template.  In particular we used the
atlas                  of                 FLAMES                 solar
spectra\footnote{http://www.eso.org/observing/dfo/quality/GIRAFFE/\\pipeline/solar.html}.
This  set of  measurements completely  covers all  high-resolution and
low-resolution settings of  the FLAMES fiber system.  We  took care to
select the spectra corresponding to our scientific setups as described
in the previus Section.  For each given instrument and setting we also
construct a  model for the  telluric contamination.  This was  done by
using the  ESO sky  correction tools and  in particular the  Sky Model
Calculator\footnote{http://www.eso.org/observing/etc/bin/gen/\\form?INS.MODE=swspectr+INS.NAME=SKYCALC}.
The model  spectra were  analyzed and the  regions with  the strongest
contamination were  then excluded from  the analysis of  the scientific
spectra.   For the  cross-correlation  analysis we  developed our  own
tools.   First an  accurate model  of  the continuum  was created,  by
averaging   the   spectra   in   20  spectral   subregions,   strongly
downweighting  absorption and  emission features.   Then  a five-order
polynomial was fit  and used to normalize the  spectrum. We calculated
the  cross-correlation   in  between  -200   km$\rm\,s^{-1}$  and  200
km$\rm\,s^{-1}$ in steps of  1 km$\rm\,s^{-1}$.  The centroid position
was calculated adopting a gaussian  function using the upper 30$\%$ of
the  CCF  measured from  the  peak.  From  the  CCF  function we  also
calculated the  bisector as described  in Queloz et al.   (2001).  The
obtained  radial  velocities  were   corrected  to  put  them  in  the
heliocentric  reference system  and the  heliocentric julian  dates at
mid-exposure were  calculated with the  task $rvcorrect$ of  IRAF.  We
then applied a quality cut, dropping from the list of measurements all
those  for  which  the CCF  peak  was  found  lower  than 0.3  or  the
associated   error  in  the   radial  velocity   was  larger   than  5
km$\rm\,s^{-1}$ or for which the  resulting spectra had a very low S/N
(tipically  lower than  3).  A  sample of  5334 GIRAFFE  and  445 UVES
measurements  remained  after these  steps  corresponding  to a  total
number of 313 stars.  All the stars monitored with UVES were monitored
also with GIRAFFE.  The median  number of measurements per star in the
final catalog is equal to 15. Fig.\ref{fig:sn}, presents the S/N
of the spectra. The S/N was calculated with the DER\_SNR algorithm
presented in Stoehr et al. (2008). Using the ESO exposure time calculator
we found a reasonable agreement between the observed S/N and the expected one.

\subsection{Systematics analysis}

\subsubsection{GIRAFFE}

\begin{figure*}
\centering
\includegraphics[width=5cm]{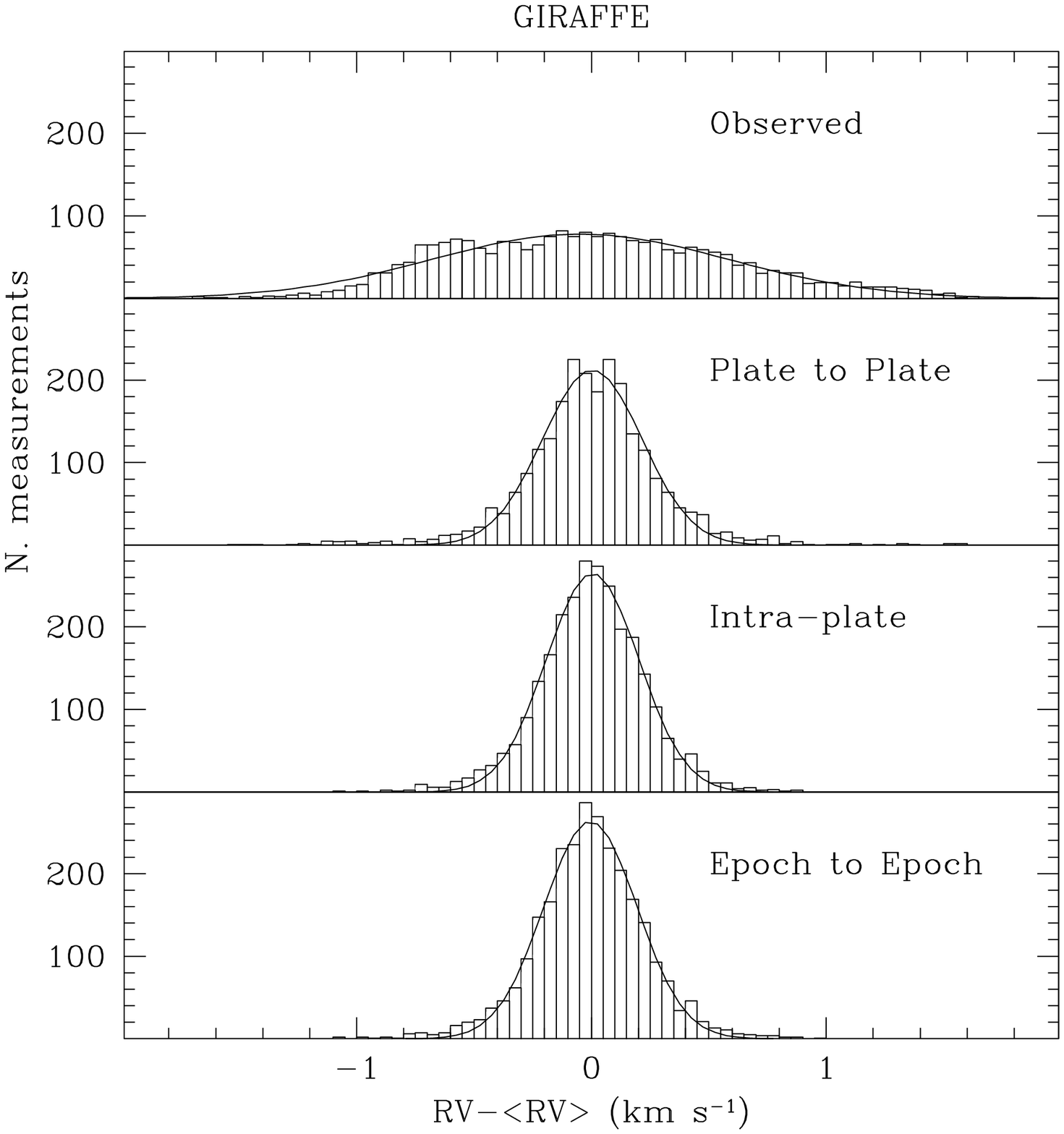}
\includegraphics[width=5cm]{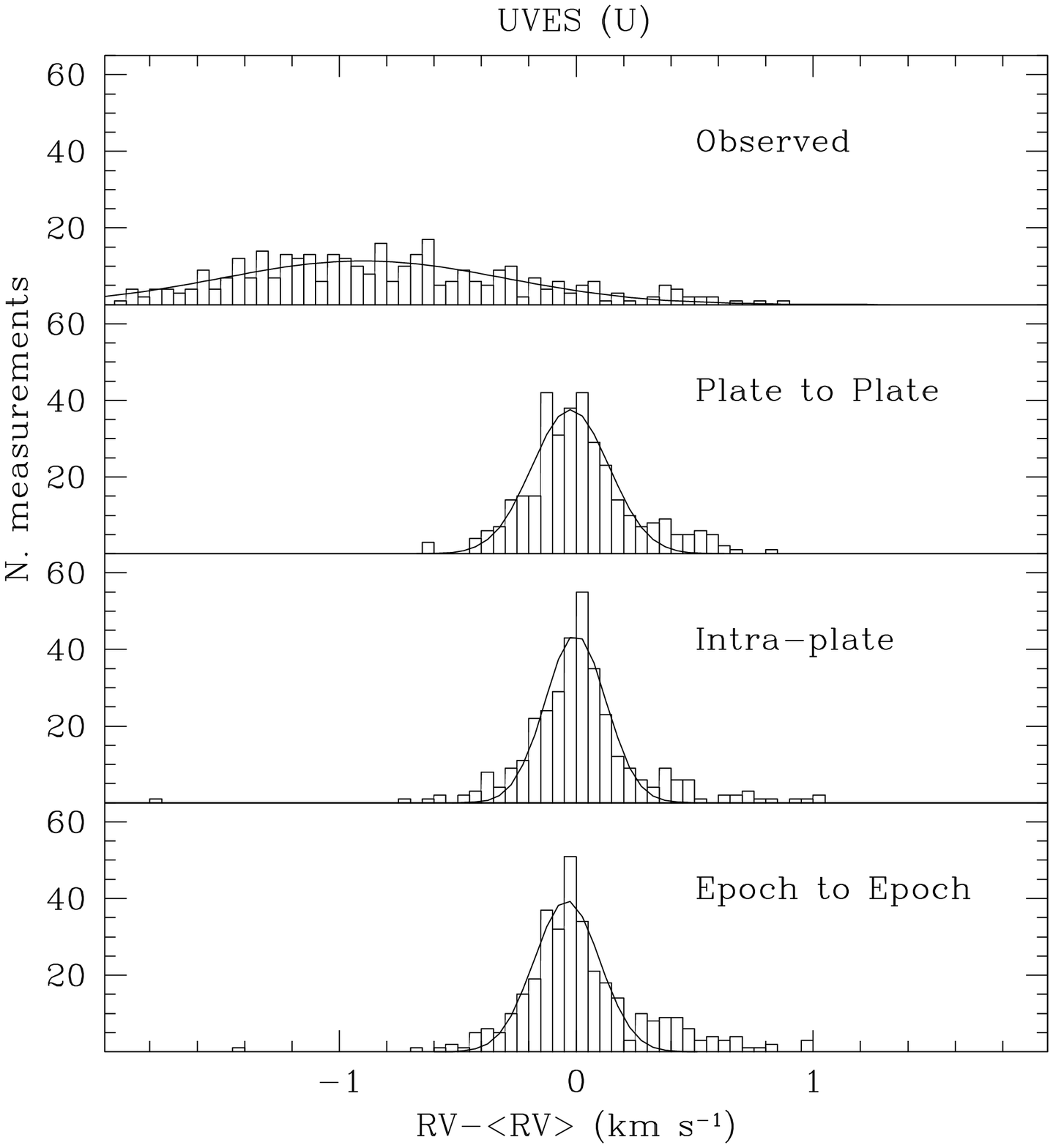}
\includegraphics[width=5cm]{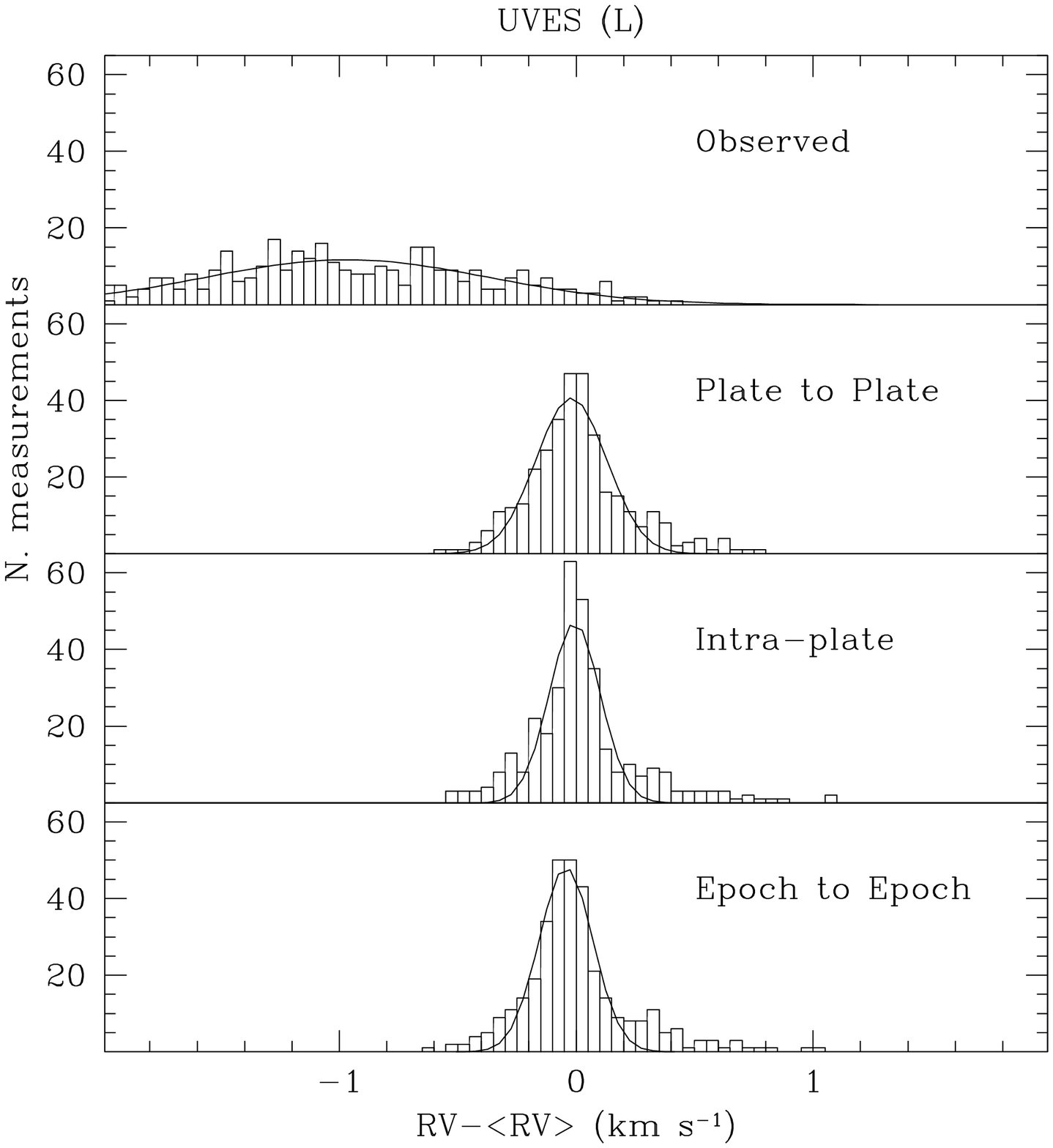}
\caption{
Average   subtracted   radial   velocities   for  GIRAFFE   and   UVES
spectrographs during  different stages of  post-processing as reported
in the panels and in the text. The continuous line represents
the best-fit Gaussian to the residual distributions.
\label{fig:epoch_zpoints_giraffe_uves}
}
\end{figure*}

\noindent
Before  merging the  datasets we  checked and  correct  for systematic
effects. First we analyzed the  GIRAFFE dataset.  We selected a sample
of reference  stars.  To do that  we imposed an RMS  treshold limit in
radial  velocity  RMS$<1$  km$\rm\,s^{-1}$.   We further  limited  the
sample to only stars with at least 10 measurements resulting in a list
of 161 stars.   We checked then for plate to  plate zero point offsets
in  radial  velocities as  illustrated  in Fig.\ref{fig:offsets}  (top
panel).   This   Figure  represents  the   average  subtracted  radial
velocities (RVs) for all  reference star GIRAFFE measurements, plotted
as  a function of  the plate  number.  A  clear pattern  of systematic
variations is present which amounts to maximum shifts of up to $\sim$1
km $\rm s^{-1}$.   To correct for this effect,  we calculated a robust
average of the  residual RVs for each plate  and subtracted this value
for     all     the     measurements     of    that     plate.      In
Fig.\ref{fig:epoch_zpoints_giraffe_uves}  (left  panel)  we  show  the
histograms  of the  residuals  RVs. We fit a Gaussian function to
these distributions to judge the improvement of the residual scatter
during the different steps of post-processing.  
In the  top  panel the  observed
residuals (before  any correction) are  visible.  The dispersion  
of the best-fit Gaussian is equal to 631 m s$^{-1}$. In the second  
panel from the
top the  effect of the plate  to plate correction is  visible.  Such a
correction substantially improve the precision of the measurements 
bringing the dispersion to 212 m  s$^{-1}$ and it is therefore of paramount
importance. The fiber  system, however, may be subjected  also to more
subtle mechanical  drifts which  may depend on  a variety  of factors.
Such systematic drifts may in turn induce intra-plate systematics.  To
check for  that we decorrelate  the radial velocity  residuals against
the fiber positioning  number. The result of this  correction is shown
in      the      third       panel      from      the      top      in
Fig.\ref{fig:epoch_zpoints_giraffe_uves}.  The corresponding dispersion
improved further being equal to 197 m s$^{-1}$  after  this   
correction  demostrating   that  intra-plate
systematics  are indeed  present, albeit  their magnitude  seem  to be
smaller than the plate to  plate systematics.  It is possible that the
resulting  radial  velocities  are   affected  also  by    variable
observing conditions during  each night.  We checked for  this effect by
regrouping all measurements  as a function of the  observing night and
decorrelating  the  residuals  against  airmass (calculated  for  each
star), color (derived from our  catalog) and allowing for a systematic
zero  point offset.  No   improvement  is observed  in this
case (the dispersion is equal
to 199 m s$^{-1}$).  We decided in any case to apply this
epoch-to-epoch correction  to the  GIRAFFE data. 
The distributions in Fig.\ref{fig:epoch_zpoints_giraffe_uves} appear close
to Gaussian. The distribution of the $observed$ residuals
in the top panel appears slightly asymmetric.

\begin{figure*}
\centering
\includegraphics[width=16cm]{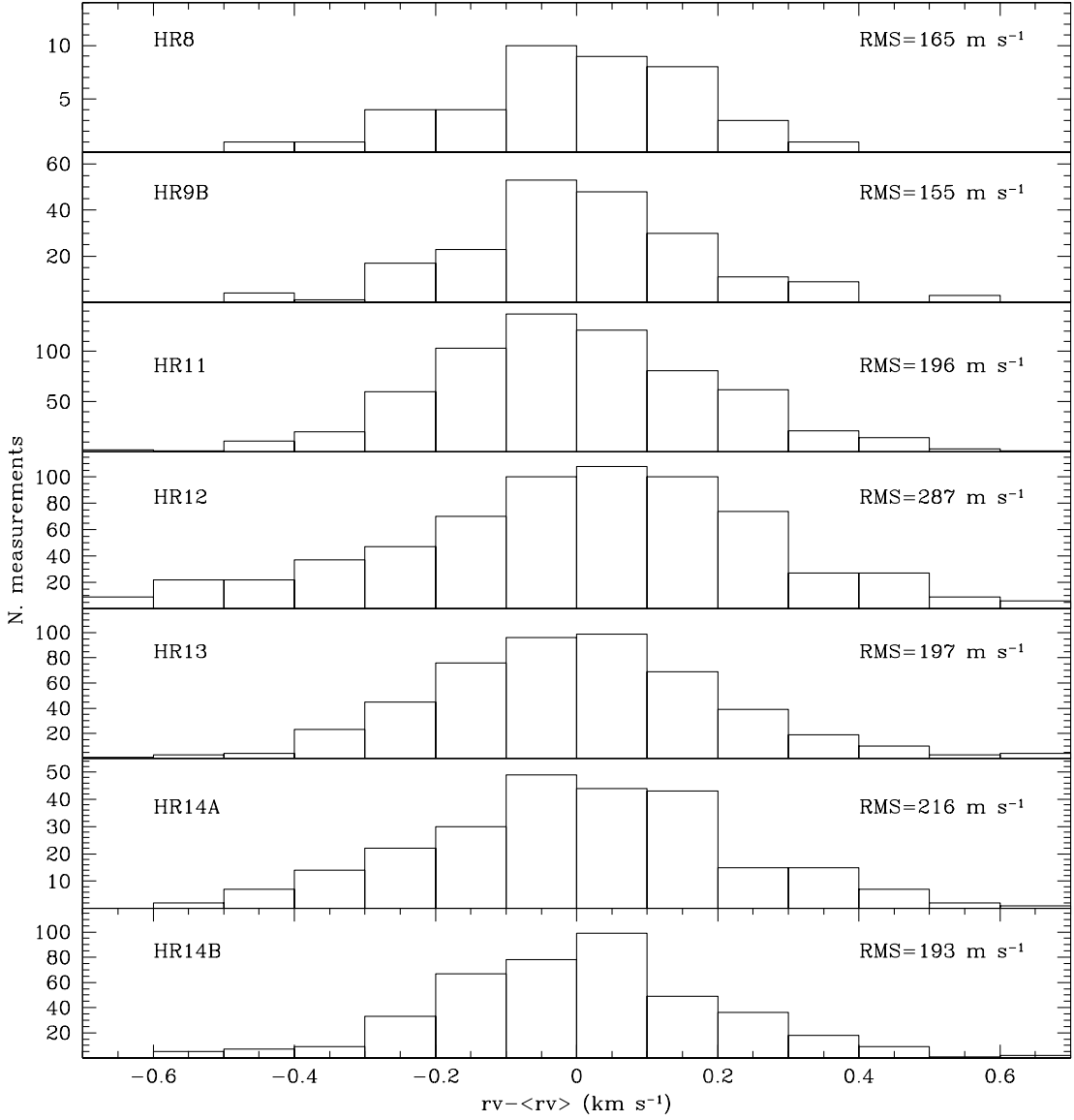}
\caption{
Radial velocity residuals for each GIRAFFE setting.
\label{fig:res_filter_giraffe}
}
\end{figure*}

Because the
GIRAFFE measurements  have been acquired  with 7 different  setups, we
checked  also for  the  presence of  setup-dependent systematics.   In
Fig.\ref{fig:res_filter_giraffe}, we regrouped all the residual radial
velocities as a  function of the used setting.   No significant offset
is observed after the correction steps reported above. However, we did
note the fact  that the different setups do not  appear to achieve the
same  performances. The  setting HR09B  delivered the  highest precision
(155 m$\rm\,s^{-1}$ as measured from  the RMS of the residuals), while
HR12 was  found the lowest  performant (287 m$\rm\,s^{-1}$).   This is
likely  due to  the strongest  telluric contamination  present  in the
spectral window  of this setting.  It appears therefore that  beyond the
systematics discussed  above, the choice of the  instrumental setup is
of great importance to achieve the highest radial velocity precisions.

The  internal error  of  the  individual  measurements has  been
initially assigned  by our  Gaussian fitting algorithm  once performing
the CCF analysis, being the error of the centroid position obtained by
the   Levemberg-Marquardt  algorithm.    After   the  correction   for
systematics  we then calculated  the ratio  of the  median RMS  to the
median internal error for UVES and GIRAFFE separately.  The individual
errors were then  multiplied by this factor.  This  approach allows to
match  the  average  error  levels  based on  the  RMS  analysis,  but
preserves the relative errors among different measurements.  

\noindent
It is interesting to compare  the results obtained by using the plates
which  were  observed  with   simultaneous  calibrators  or  not.   In
particular the plates without calibrators were those obtained with the
HR14B setting.   A look at  Fig.\ref{fig:res_filter_giraffe} indicates
that the final RMS obtained for  this setting is in general not worser
the  one  obtained with  other  settings.   By  comparing instead  the
distribution  of the observed  residuals (without  any post-correction
applied, therefore  the result of  the pipeline) we obtained  for this
setting a  dispersion equal to 690  m s$^{-1}$, worser than  the 631 m
s$^{-1}$ quoted  above for the  entire sample.  Using only  the plates
with calibrators we obtain 632 m  s$^{-1}$, very close to the value of
the entire  sample, because plates with calibrators  are dominant.  It
results therefore that the correction applied by the pipeline based on
the  simultaneous  calibrators  does  not  improve  substantially  the
precision,  especially in  comparison with  the final  values  that we
obtained by means  of our analysis.  In a recent  work Malavota et al.
(2015) illustrated  that the  wavelength calibration  solution  may be
inaccurate  in the  first  place due  to  the presence  of non  linear
distorsion  terms  between  the  wavelength solution  of  the  morning
calibration  and  the  simultaneous  calibrations. This  in  turn  can
produce artificial zero points offsets.   It may be also possible that
the  simultaneous calibrators  are not  tracking perfectly  the radial
velocity zero  points offsets  due to the  presence of  a differential
offset between  the calibrators and the  scientific fibers. Inaccurate
fiber  scrambling could be  a possible  reason.  If  the light  is not
uniformly spread across the fiber, spurious radial velocity variations
can  be produced,  as  a  consequence of  thermal,  pressure or  other
environmental variables (e. g. seeing) variations.  Our analysis seems
to indicate that the plate-to-plate systematics are the most prominent
sources of noise.  

\subsubsection{UVES}

\begin{figure*}
\centering
\includegraphics[width=14cm]{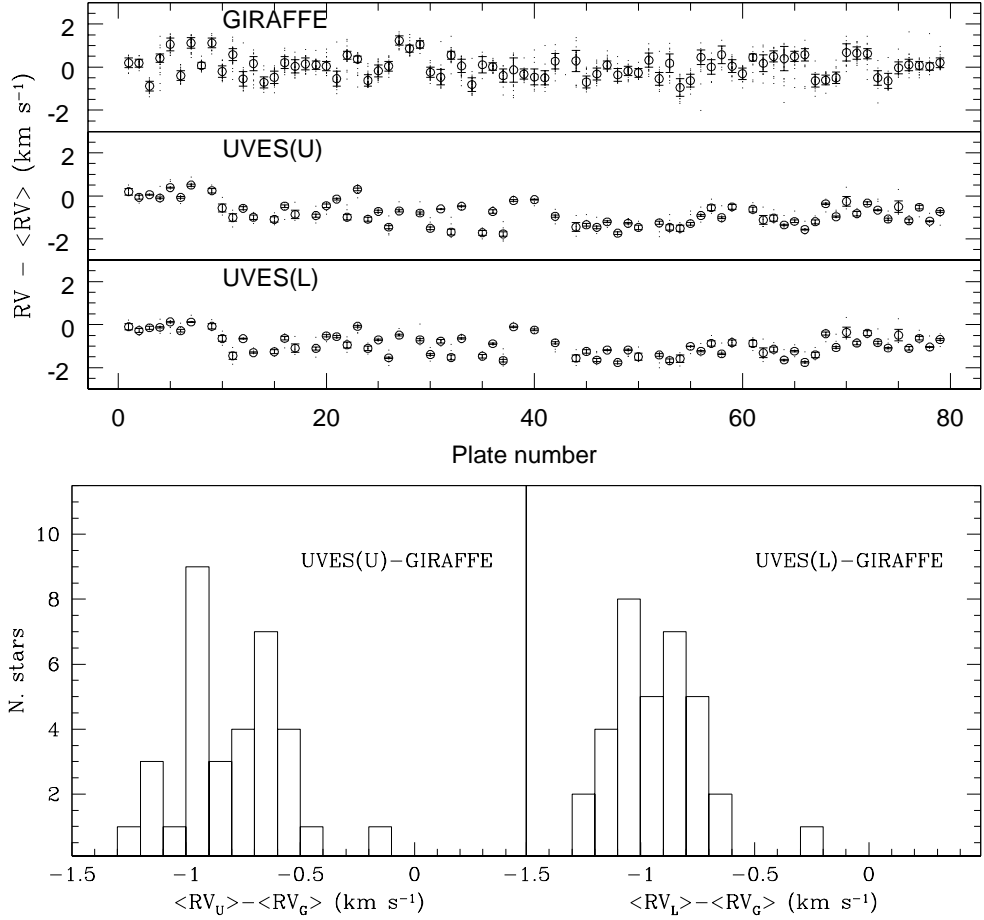}
\caption{ Plate  to plate radial velocity systematics  for GIRAFFE and
UVES as a function of plate  number (top panels).  In the bottom panel
we  report the  histogram of the  difference between  the average
radial velocity  calculated in  the GIRAFFE reference  system (<RV$\rm
_G$>) and in the UVES  reference system (<RV$\rm _U$> and <RV$\rm _L$>
for  the upper  and lower  UVES spectra  respectively).   Only GIRAFFE
reference stars  having also 5  UVES radial velocity  measurements (34
stars) are considered.  
\label{fig:offsets}
}
\end{figure*}

\noindent
The FLAMES-UVES detector is the  mosaic of two different CCDs covering
the bluer and redder portion of  the spectral range of this setup. The
pipeline provides an output for each  one of them and we analyzed them
independently. In the following we will indicate them as the lower (L)
and  upper (U)  UVES spectra.   The  average value  of the  systematic
corrected  GIRAFFE measurements  was  adopted as  the radial  velocity
reference  system to  correct the  UVES measurements.   With  a median
value  of 15  measurements per  star with  an RMS  down to  around 200
m$\rm\,s^{-1}$, and  an overlap of  3-4 references per UVES  plate the
precision of such  a reference system can be expected  at the level of
20    m$\rm\,s^{-1}$.     Fig.\ref{fig:offsets},    illustrates    the
plate-to-plate systematics  of the  UVES spectrograph with  respect to
such  a reference  system.  Once  compared with  the  GIRAFFE observed
residuals (illustrated in the same figure) it appears that the the two
spectrographs  did not follow  the same  pattern of  systematics.  The
upper and  lower UVES spectra delivered instead  very similar results.
Importantly, we can clearly see  the presence of an offset between the
radial velocity  scale of  the two spectrographs.   
We selected
then the stars  in common between the UVES  and GIRAFFE datasets which
were considered GIRAFFE reference  stars (as defined above) and having
also at least  5 UVES measurements. By using this  sample (in total 34
stars)  we  calculated, for  each  star,  the  difference between  the
average radial velocity calculated using only GIRAFFE measurements and
only  UVES  measurements (this  time  considered in the UVES  reference
system).  

The histogram of these differences are illustrated in the
bottom panels  of  Fig.\ref{fig:offsets}.   Both the upper  and lower
spectra  of UVES  deliver  offset radial  velocities  with respect  to
GIRAFFE.  The average offset is found equal to:

\begin{equation}
\rm \overline{RV}_{UVES}-\rm \overline{RV}_{GIRAFFE}=(-0.87\pm0.15)\rm\,km\,\rm s^{-1}.
\end{equation}

\noindent
Subsequently  we  compared  the  radial  velocity  scale  of  the  two
spectrographs  with the  one  of HARPS  (for  the stars  in common  as
reported below),  and we concluded  that the measurements  in the
reference system of  GIRAFFE are in the  reference system of HARPS
with no noticable offset.

\noindent
Fig.\ref{fig:epoch_zpoints_giraffe_uves} reports  the results obtained
applying the plate-to-plate correction  to UVES data (middle and right
panels).  Also  in this case  we observe a substantial  improvement of
the precision.  The dispersion of the best  Gaussian fit to the
observed distribution is  in this case 598 m  s$^{-1}$ (average of the
upper  and  lower  UVES   spectra),  while  after  the  plate-to-plate
correction the dispersion is equal to 156 m s$^{-1}$.   Thanks to the
adoption of  the GIRAFFE reference system  we were then  able to check
also  for  intra-plate  systematics   for  UVES.   Note  that  such  a
correction is  not possible using only the  simultaneous fibers, given
that only one simultaneous fiber per plate is allowed for UVES.  
The adoption of this correction improved further the dispersion (119 m
s$^{-1}$)   although  we   note  that   the   corresponding  residuals
distribution presents evident  non-Gaussian tails.  Likely 
intra-plate  systematics are present although  the smaller number
of stars used for the correction  may not be able to perfectly capture
the  systematic  trend.  .  We  proceeded  further  and analyzed  the
epoch-to-epoch UVES systematics,  decorellating all residuals acquired
during the  same observing night as  a function of  color, airmass and
allowing for the presence of a zero point offset.  In this case such a
correction  appears to  deteriorate (127  m s$^{-1}$)  the
precisions, and we therefore did not apply it.

\noindent
After the  above reported correction  steps, the upper and  lower UVES
spectra  were merged, and  the results  finally merged  to the  one of
GIRAFFE to produce a single  combined catalog. For what has been said,
the radial velocity scale of this  catalog is the one of GIRAFFE (that
is  the one  of HARPS,  see below).   The GIRAFFE  and combined
catalogs contain 313  stars in total, while the  UVES catalog contains
75 stars.  The total number of stars with at least two measurements is
equal to  300 stars for  GIRAFFE (and combined catalog) and 67  stars 
for UVES.

\begin{figure*}
\centering
\includegraphics[width=16cm]{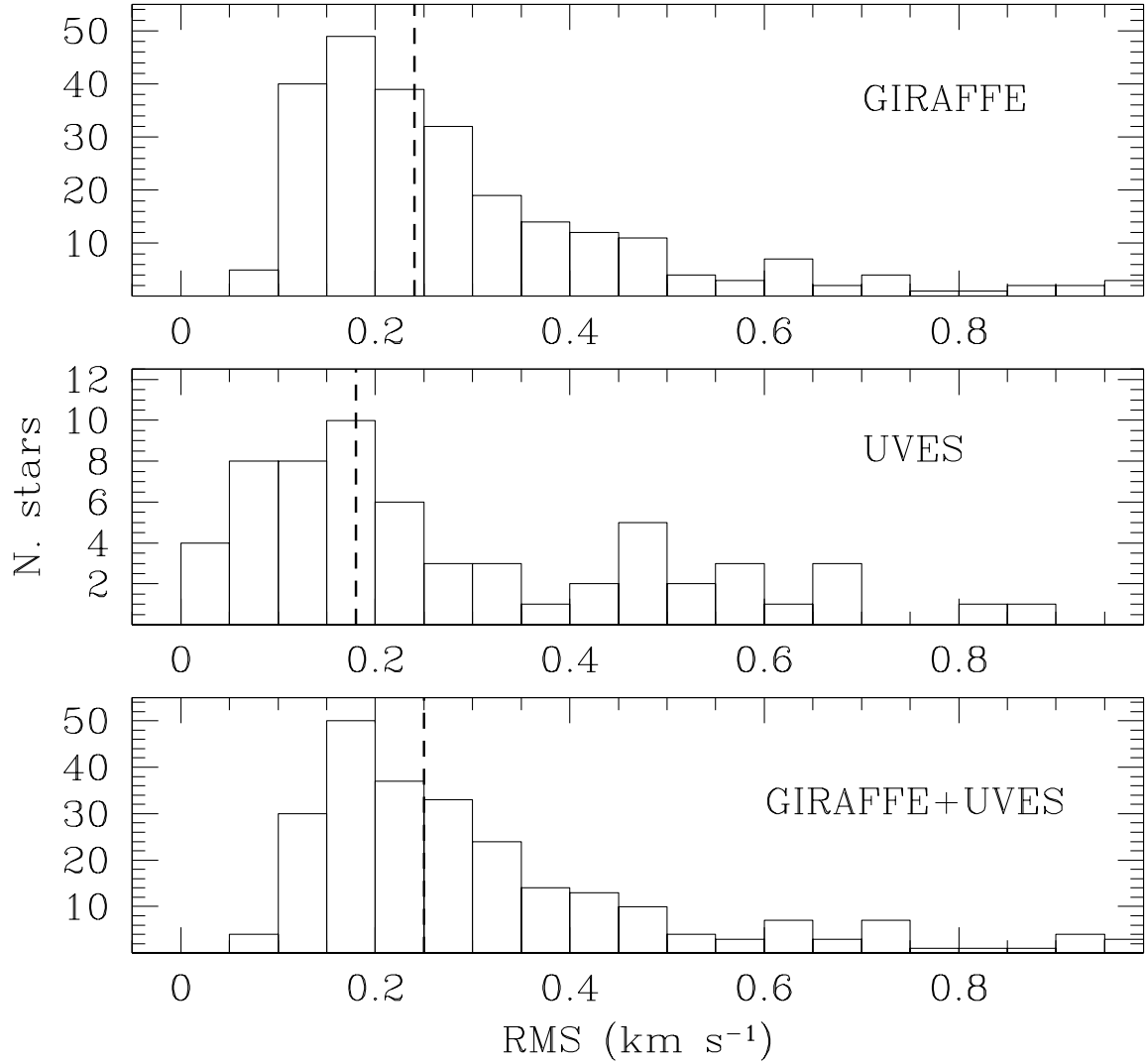}
\caption{
Top: histogram  of radial  velocity corrected RMS  
for  all stars with at least two measurements and RMS below 1 km s$^{-1}$ 
observed  with
GIRAFFE (top), UVES (middle) and in the combined catalog (bottom).
The dashed vertical lines denote median values.
\label{fig:rms_hist}
}
\end{figure*}

\noindent
Fig.\ref{fig:rms_hist} shows  the final RMS obtained  for the GIRAFFE,
UVES  and combined  sample for  all the  stars 
having  at  least two
measurements  respectively and with RMS<1 km$\rm s^{-1}$
(250 stars for GIRAFFE and combined catalogs and 61 stars for UVES catalog).  

\begin{figure*}
\centering
\includegraphics[width=16cm]{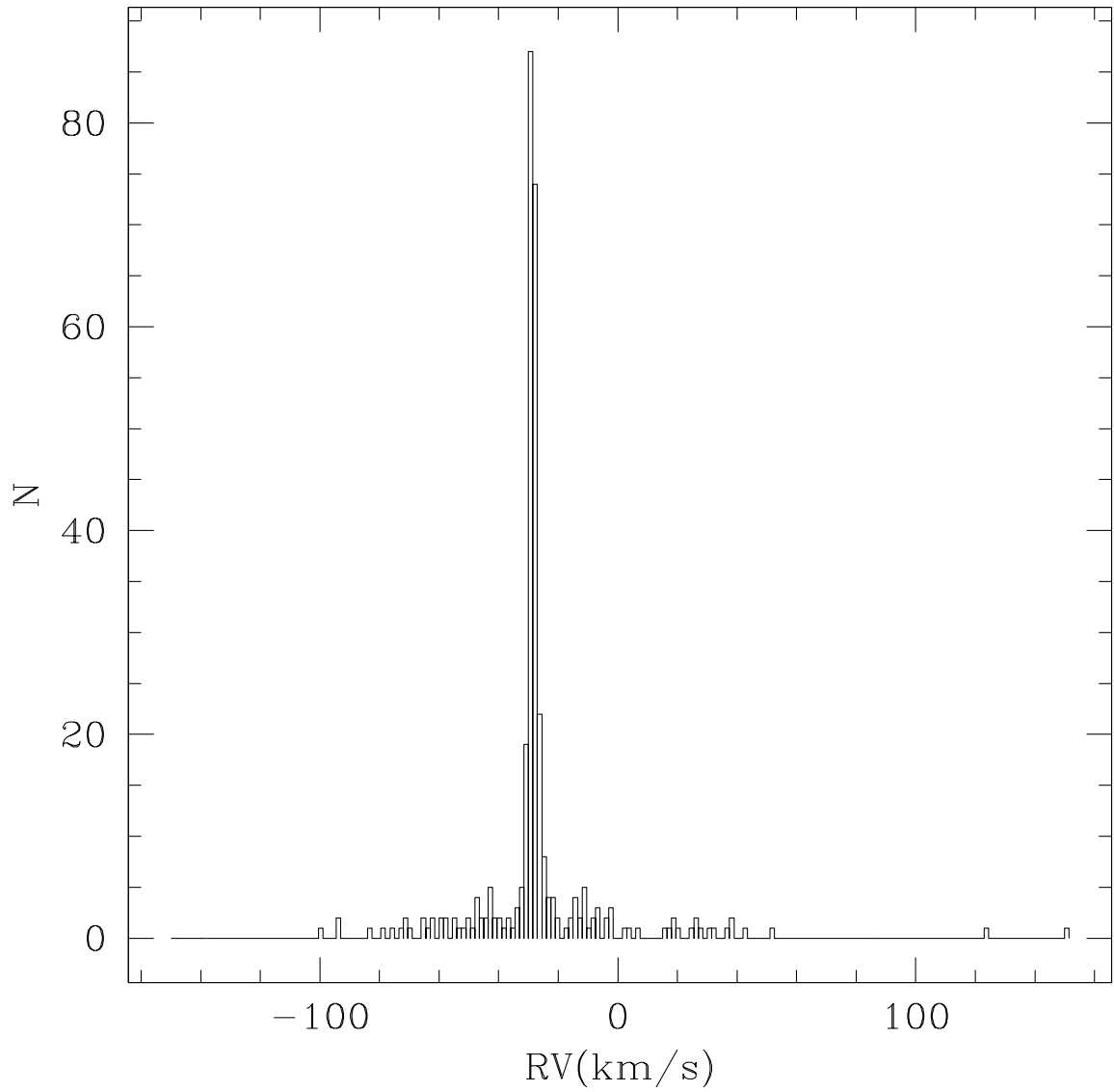}
\caption{
Histogram  of all  radial  velocities measurements  acquired with  the
FLAMES spectrograph.
\label{fig:rv_hist}
}
\end{figure*}

The median  precision (denoted by the vertical dashed lines 
in Fig.\ref{fig:rms_hist}) 
of GIRAFFE is  240 m $\rm  s^{-1}$ while  the one of  UVES is 180  m $\rm
s^{-1}$.  The combined sample median RMS (250 m $\rm s^{-1}$) reflects
the GIRAFFE sample  RMS due to the much  larger number of measurements
acquired   with   GIRAFFE.    Fig.\ref{fig:rv_hist},  represents   the
histogram  of  all  the  radial  velocity measurements  in  the  final
catalog.  The peak due to cluster members is clearly visible.

\subsubsection{HARPS}

\noindent
The HARPS  radial velocities measurements were  obtained directly from
the header of the pipeline processed files, and no post-correction was
applied.  This  sample is composed by a  set of 7 stars  that were all
included  in the  UVES-GIRAFFE dataset  observed in  2004-2005.  While
four of  these objects appeared  possible planetary candidates  at the
time of  the follow-up, the  HARPS observations have dismissed  all of
them as false positives.

\noindent
On the contrary, we note that the transiting planetary candidate (star
171895) we found  in our previous photometric  campaigns, described in
M11,  has not  been targeted  with HARPS  and it  remains  therefore a
primary target for future observational efforts.

\noindent
To estimate the average precision  obtained by HARPS on this sample of
stars we subtracted the mean radial velocity and calculated the RMS of
all  combined residuals.  In this  way  we obtained  a value  of 35  m
$\rm\,s^{-1}$.

\noindent
By comparing the average velocity obtained with GIRAFFE and with HARPS
for  these stars  we obtained  an average  difference between  the two
spectrographs radial velocity scale equal to (-0.094$\pm$0.091) km$\rm
\,s^{-1}$, which demonstrates that  the radial velocity scale of GIRAFFE
is consistent with the one of HARPS.

\section{Search for variables}
\label{sec:search}

\begin{figure*}
\centering
\includegraphics[width=16cm]{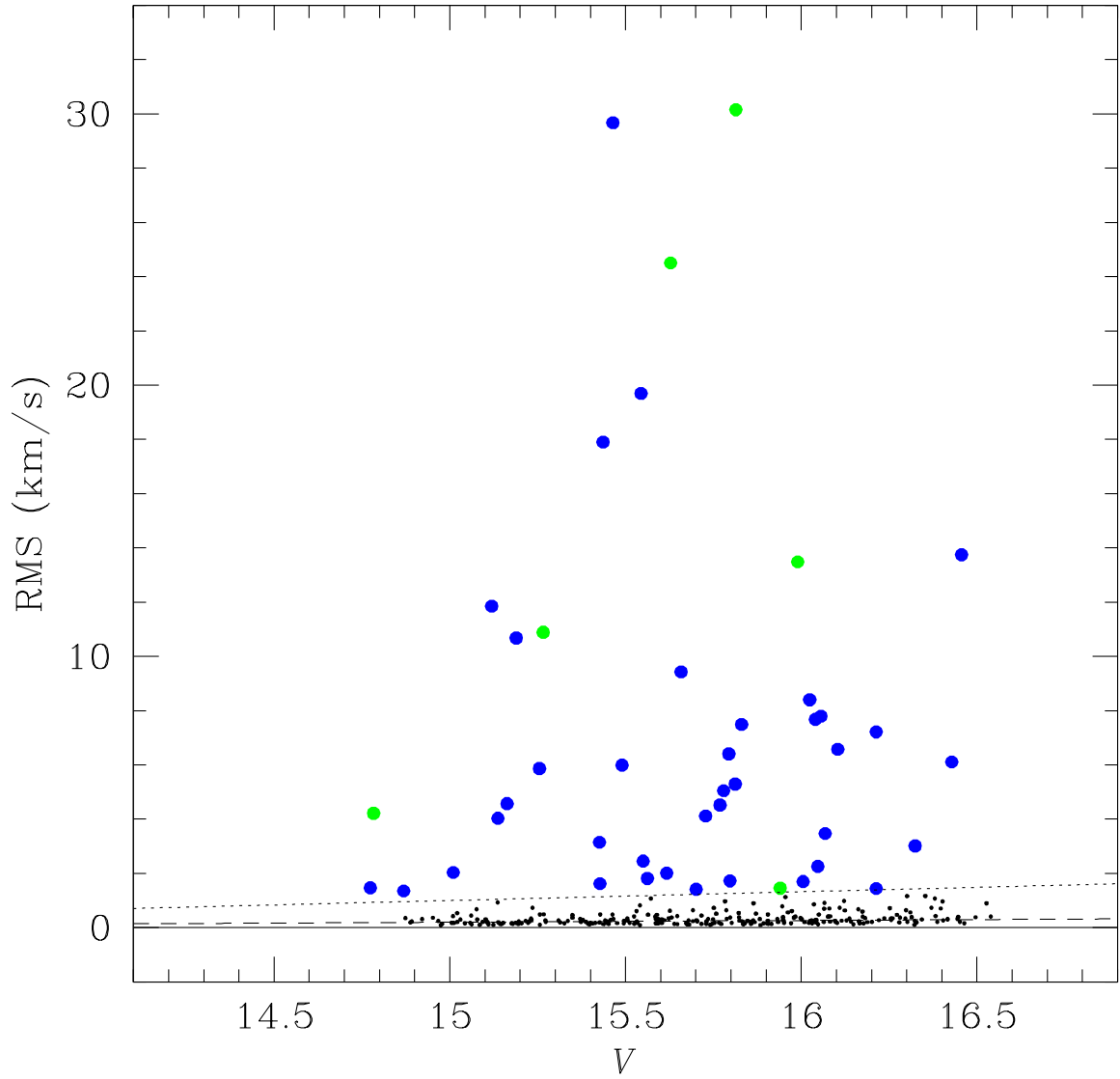}
\caption{
Radial velocity  RMS for  all stars  in our catalog  as a  function of
magnitude.   The  dashed  line  denotes the  magnitude  dependent  RMS
treshold  described   in  the  text,  the  dotted   line  denotes  the
5$\rm\sigma$  detection treshold.  Blue points  are RMS  variables and
green points are GLS variables.
\label{fig:RMS}
}
\end{figure*}

\begin{figure*}
\centering
\includegraphics[width=16cm]{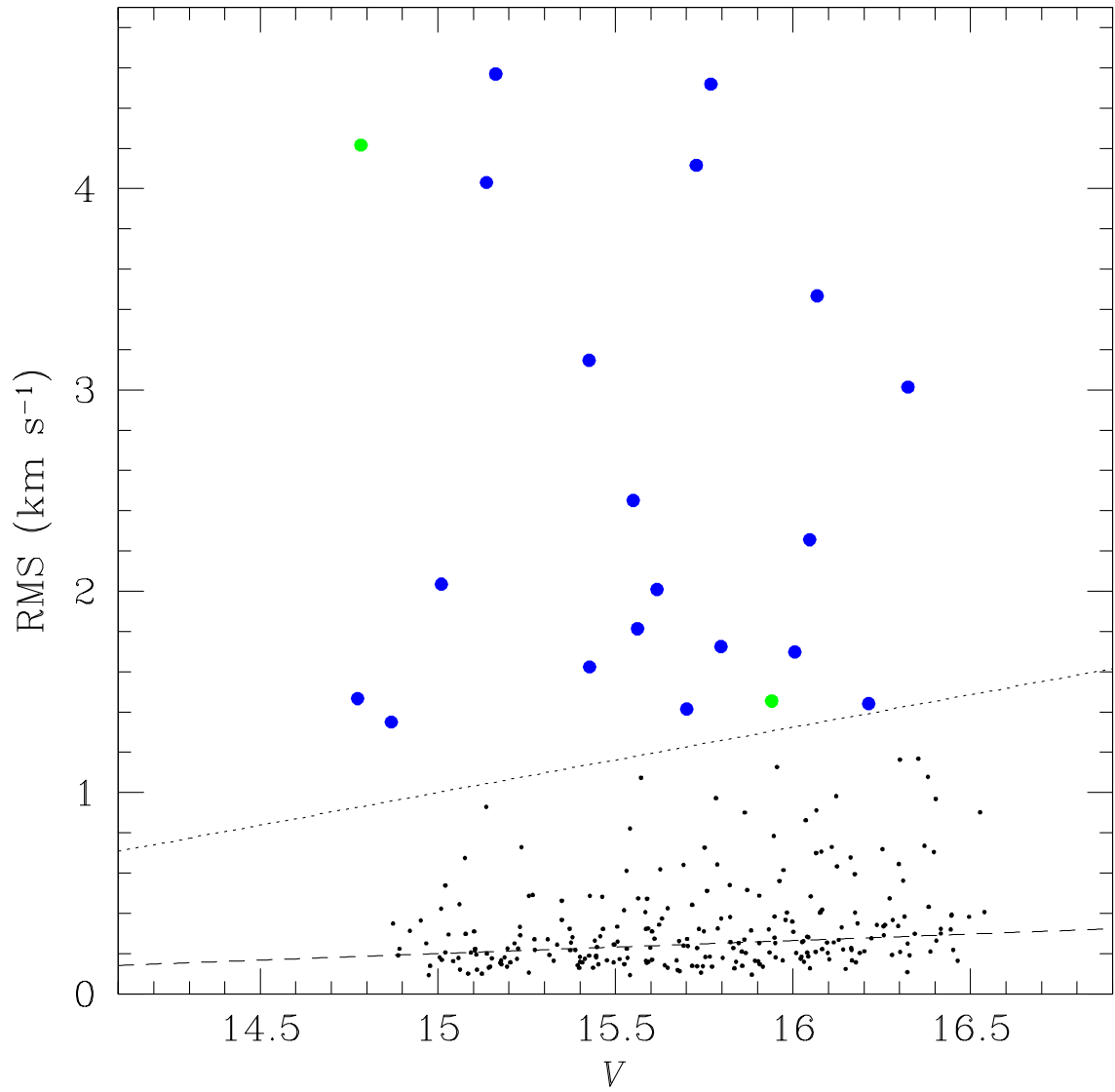}
\caption{
Radial velocity RMS for all stars in our catalog in the region below 5
km $\rm\,s^{-1}$.
\label{fig:RMS_CLOSE}
}
\end{figure*}

Figure~\ref{fig:RMS} shows  the RMS  distribution for all  the catalog
stars.  We adopted a magnitude dependent treshold for the detection of
variables  as visible in  Fig.~\ref{fig:RMS_CLOSE}. This  RMS treshold
($\rm \sigma_{ref}$) was calculated by medianing the RMS values of the
stars  found  in 0.2  mag  bins  and fitting  a  linear  model to  the
resulting median  values.  We  adopted a 5$\rm  \sigma_{ref}$ treshold
denoted  by the  dotted  line  in the  above  mentioned Figures.   All
objects beyond  this treshold were considered  bona-fide variables and
are denoted in particular by the blue dots.

\noindent
We also searched for periodic variables given the extent of our survey
and  the fact  that many  objects  have multiple  measurements in  our
catalog.   In particular we  limited the  search for  periodicities to
stars with at  least 10 epochs.  We used  the Generalized Lomb Scargle
(GLS) algorithm  of Zechmeister \& K\=urster (2009)  and in particular
their  Eq.~5.  This algorithm  allows to  fit a  sinusoid to  the data
together  with  a constant  offset.   To  properly  set the  detection
treshold  of  the  GLS  we  constructed  a  mock  sample  of  constant
artificial  radial velocities.   The measurements  for each  star were
simulated assuming  a gaussian noise  with a dispersion equal  to $\rm
\sigma_{ref}$.  We  considered the true distribution of  epochs of our
stars,  each simulated  star  being a  constant  mock copy  of a  real
observed one. We created around 10$^5$ simulated stars and applied the
GLS  to all  of them.   We searched  for periodic  signals considering
periods  in between  0.5 and  1000 days  subdividing this  interval in
10000  equal frequency  steps.  From  the distribution  of  GLS powers
($\rm  p_{GLS}$) we  obtained, we  imposed a  False  Alarm Probability
(FAP) treshold equal to 0.1$\%$, which was found equal to $\rm p_{GLS,
  FAP}=0.956$.

\noindent
In the  exactly same way we  applied the GLS to  all real measurements
flagging  out all  objects having  $\rm p_{GLS}\,>\,p_{GLS,  FAP}$. We
found  that this  condition  was  met for  28  objects. We  considered
however  as reliable  GLS variables  only those  for which  an orbital
solution      could      be       found      as      described      in
Sect.~\ref{sec:spectroscopic_binaries}.  Most  of  the  GLS  solutions
implied long  term trends or  in any case  the phase coverage  was not
considered  sufficient to perform  a fit.   All  of these
objects lie  beyond the  5$\rm\sigma_{ref}$ treshold and  are therefore
also RMS variables. 

The final list of reliable  GLS variables contains 6 objects which are
denoted    by   the   green    dots   in    Figure~\ref{fig:RMS}   and
Figure~\ref{fig:RMS_CLOSE}.  The  total list of  variables contains 45
stars.


\section{Classification of the stars}
\label{sec:classification}

We  proceeded  by  calculating  the  average radial  velocity  of  the
cluster.   To do  that we  used both  the proper  motion probabilities
derived in M09 and the  radial velocities calculated in this work.  We
first considered those objects  for which the proper motion membership
probability was  larger than 90$\%$  at magnitude $V=$12.5  and larger
than 50$\%$ at $V=18$,  interpolating linearly between these extremes.
Furthermore, we impose  that the corresponding x and  y coordinates of
these   stars  should   have  been   contained   within  $800<x<1300$,
$1000<y<3000$  in the  reference system  of  CCD51 of  the Wide  Field
Imager detector  where the cluster  was located in M09.   Beyond these
limits  we  considered our  proper  motions  doubtful  as reported  in
earlier works.   These tresholds are the  same adopted in  the past to
isolate cluster  members.  We  also limited the  sample to  stars that
were  not  flagged  out  as  variables  accordingly  to  the  analysis
performed in the  previous Section. In this way  we identify 66 stars.
The  average  cluster  radial  velocity  $\rm  \overline{RV}_{cl}$  we
obtained is

\begin{equation}
\rm \overline{RV}_{cl}=(-28.81\pm0.003)\,km\,\rm s^{-1}
\end{equation}

\noindent
where the  error quoted  is the error  of the  mean. The RMS  was $\rm
\sigma_{cl}$=1.05 km $\rm s^{-1}$.  To obtain this result we used 1531
individual measurements.

\begin{figure*}
\centering
\includegraphics[width=16cm]{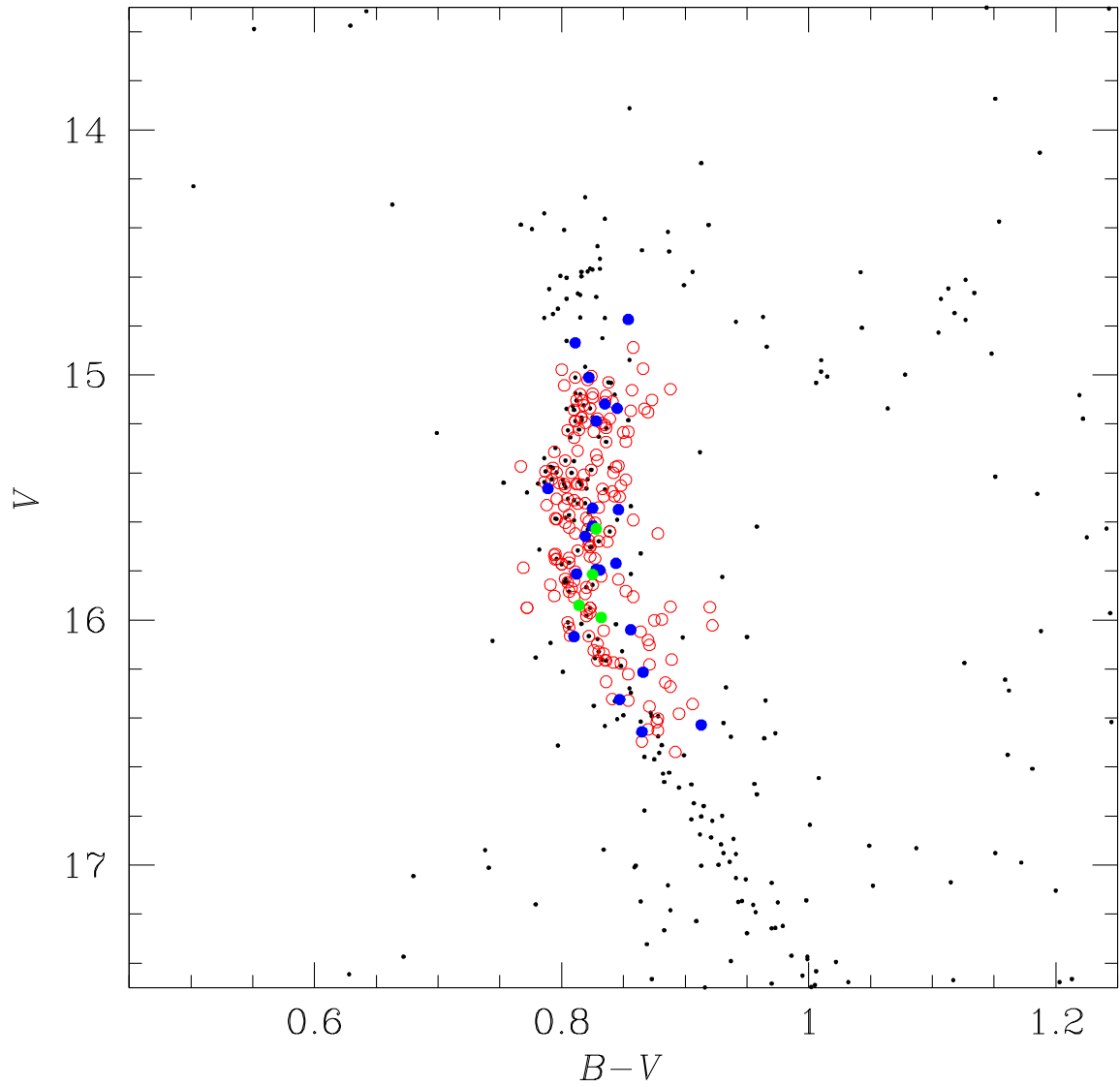}
\caption{
Color  magnitude diagram  for cluster  members.  Blue  and  green dots
represent RMS and GLS variables respectively. 
Red open circles represent radial velocity constant stars.
On  the background  the  CMD of  proper  motion members  from the  M09
catalog.
\label{fig:MEMBERS}
}
\end{figure*}

\begin{figure*}
\centering
\includegraphics[width=16cm]{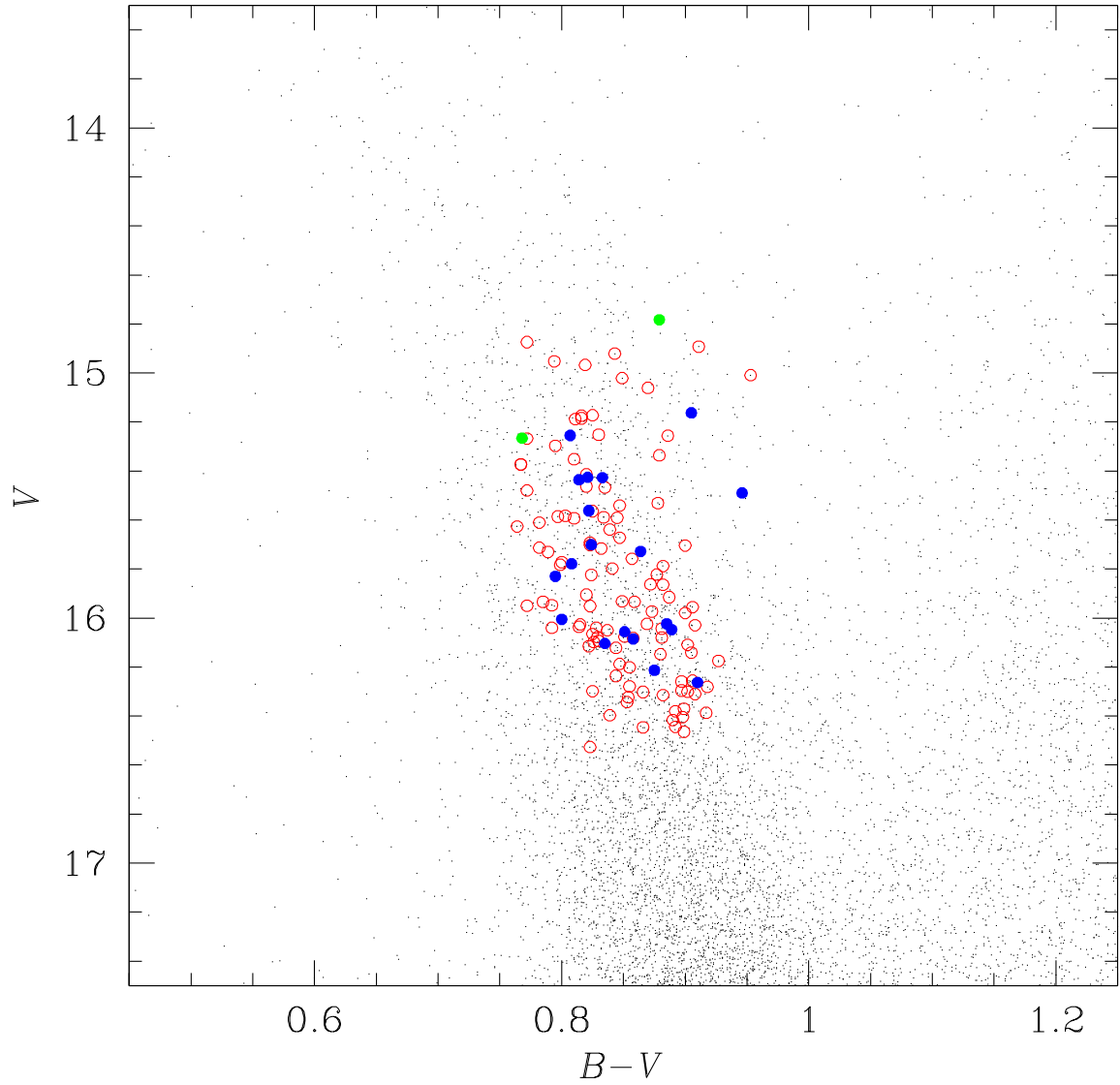}
\caption{
The same as Fig.\ref{fig:MEMBERS} for field stars.
\label{fig:FIELD}
}
\end{figure*}

\noindent
We proceeded as follows to  classify all our stars into likely cluster
members  and likely field  stars. Cluster  members were  identified as
stars which  proper motion probability was  satisfying the constraints
reported  above  and which  average  radial  velocity  was in  between
$\pm$3$\rm \sigma_{cl}$ from the cluster mean average velocity
\footnote{Note  that for stars  for which  we were  able to  derive an
  orbital solution  the systemic  velocity was considered  rather than
  the average velocity}. In absence  of proper motions (or in the case
proper motions  were considered doubtful)  we considered as  bona fide
members those stars satisfying only the radial velocity criterium. The
total sample of cluster stars is  equal to 192 stars, 25 of which were
flagged  out as  variables. Consequently,  the sample  of  field stars
contains 121  stars, 20 of which  are variables. On the  basis of this
classification we  obtain an observed frequency of  variables equal to
13$\%$ for  the cluster, and equal  to 17$\%$ for the  field.  We also
obtain  that the overall  field contamination  is around  38$\%$. Some
caveats should be kept in mind however. Among all cluster members only
59  satisfy the  stringent constraints  on proper  motions  and radial
velocity,    the   remaining   ones    are   only    radial   velocity
members. Therefore a certain  field contamination can be expected from
field    stars    sharing     the    same    velocity    of    cluster
members. Equivalently, among field  variables some fail to satisfy the
radial    velocity   criterium   but    have   high    proper   motion
probabilities. Since they are variables, their average radial velocity
is   not   necessarily   coincident   with  the   systemic   velocity.
Figure~\ref{fig:MEMBERS}  and  Figure~\ref{fig:FIELD}  show the  color
magnitude diagrams  for cluster and  field stars respectively.  In the
background  we displayed also  proper motion  likely members  from our
catalog in  the first case and  all other objects in  the second. From
these figures it is clear that  field stars are much more dispersed in
the CMD, and that cluster  members closely follow the main sequence of
the cluster. Blue  and green dots have the  same meaning than previous
figures, denoting  respectively RMS and GLS variables.   Red open dots
denote constant stars.

Table~\ref{tab:cluster_variables}  and Table~\ref{tab:field_variables}  
show the complete list of variables along with their basic parameters.

\section{Analysis of periodic variables}
\label{sec:spectroscopic_binaries}

For six variables  we were able to derive  an orbital solution.  The
period was fixed to the GLS  period corresponding to the peak with the
strongest power.  We modeled the Keplerian motion as:

\begin{equation}
\rm RV\, = \, \tilde{K} \frac{\cos u +k}{\sqrt{1-h^2-k^2}} + \gamma 
\end{equation}

\noindent
where $\tilde{K}$  is the  radial velocity semi-amplitude  without the
contribution of the eccentricity  $e$, $k=e\,\cos \omega $, $h=e\,\sin
\omega   $,  $\gamma$   is   the  barycentric   radial  velocity   and
$u=\nu+\omega$ is  the true argument  of latitude with $\nu$  the true
anomaly and $\omega$ the argument of the pericenter.

\noindent
We considered as  free parameters: $\tilde{K}$, $h$, $k$,  the time of
maximum  radial velocity  ($t_{MAX}$) and  $\gamma$.   The convergence
toward   the  best-fit   solution   was  obtained   by   means  of   a
Levenberg-Marquardt algorithm (Press  1992), and the uncertainties and
the best fit values of the parameters by means of a Markov Chain Monte
Carlo analysis.  For  each radial velocity curve we  created 20 chains
of 10$^{5}$ steps.  Each chain is started from a point 5-$\sigma$ away
(in  one  randomly  selected  free parameter)  from  the  best-fitting
solution   obtained  by   the   Levenberg-Marquardt  algorithm.    The
$\chi^2_{\rm old}$ of the fit of this initial solution is recorded and
compared  with  the  $\chi^2_{\rm  new}$  obtained  in  the  following
step. The following step is obtained jumping from the initial position
to  another  one  in  the multidimensional  parameter  space  randomly
selecting  one of the  free parameters  and changing  its value  by an
arbitrary  amount  which is  dependent  on  a  jump constant  and  the
uncertainty $\sigma$  of the parameter itself.  Steps  are accepted or
rejected  accordingly   to  the  Metropolis-Hastings   criterium.   If
$\chi^2_{\rm  new}$  is lower  than  $\chi^2_{\rm  old}$  the step  is
executed,      otherwise     the     execution      probability     is
$P=e^{-\Delta\chi^2/2}$         where        $\Delta\chi^2=\chi^2_{\rm
  new}-\chi^2_{\rm old}$.   In this  latter situation a  random number
between 0 and 1 is  drawn from a uniform probability distribution.  If
this number is lower than $P$ then the step is executed, otherwise the
step  is rejected and  the previous  step is  repeated instead  in the
chain.  In  any case  the value of  the $\chi^2$  of the last  step is
recorded and compared with the one of the following step up to the end
of the chain.  We adjusted the jump constants (one for each parameter)
in such a way that the step acceptance rate for all the parameters was
around 25 per  cent.  We then excluded the first 20  per cent steps of
each chain to avoid the  initial burn-in phase, and for each parameter
we merged the remaining part  of the chains together.  Then we derived
the  mode  of the  resulting  distributions,  and  the 68.3  per  cent
confidence limits  defined by the 15.85th and  the 84.15th percentiles
in the cumulative distributions.

\begin{figure*}
\centering
\includegraphics[width=16cm]{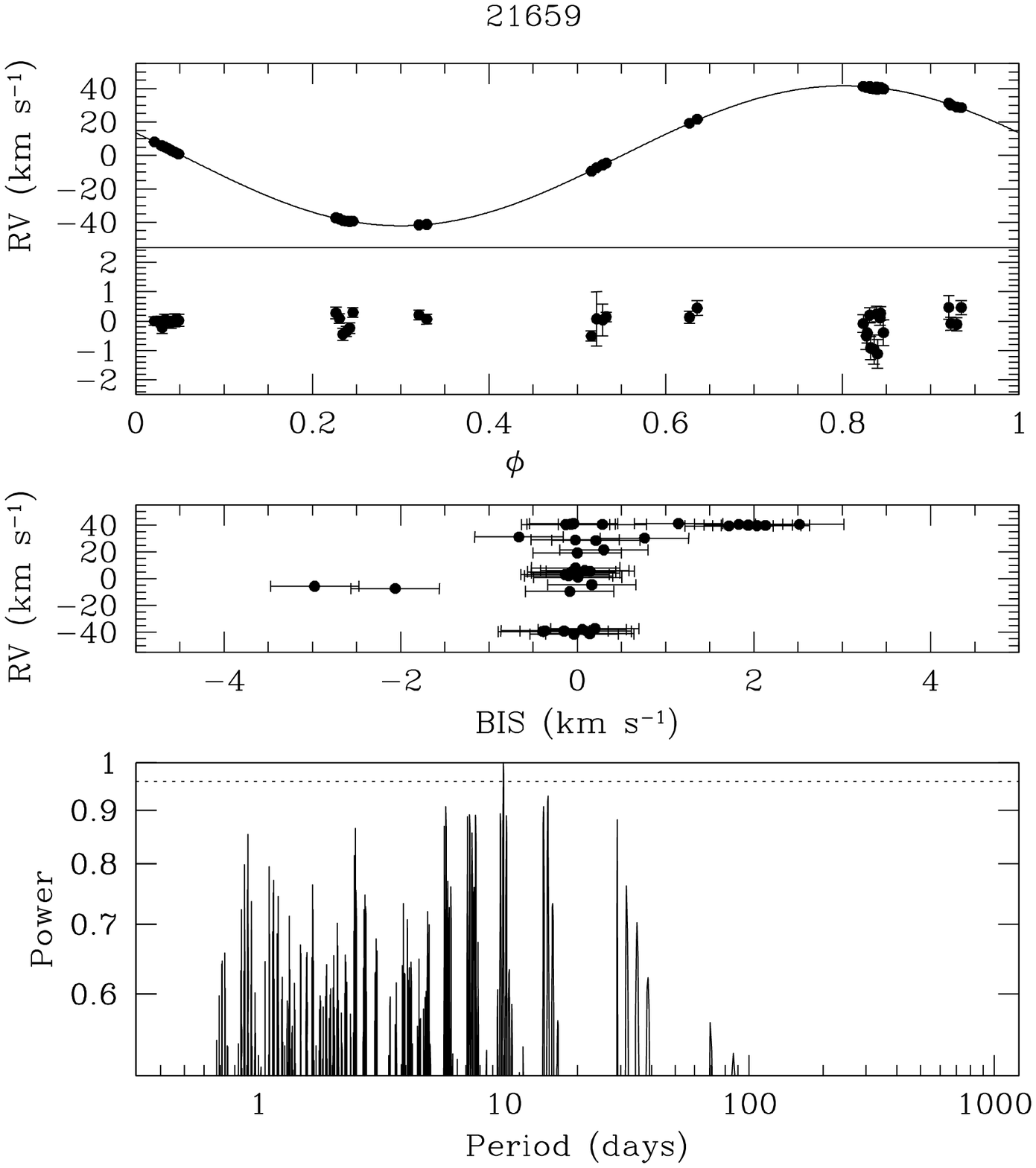}
\caption{
Folded  RV  curve with best fit  model  and 
residuals  (top), bisector
(middel) and periodogram (bottom) for star 21659.
The dashed horizontal line in the periodogram
panel denotes the GLS algorithm detection treshold
(p$\rm_{GLS,FAP}$) as defined in the text.
\label{fig:v21659}
}
\end{figure*}

\begin{figure*}
\centering
\includegraphics[width=16cm]{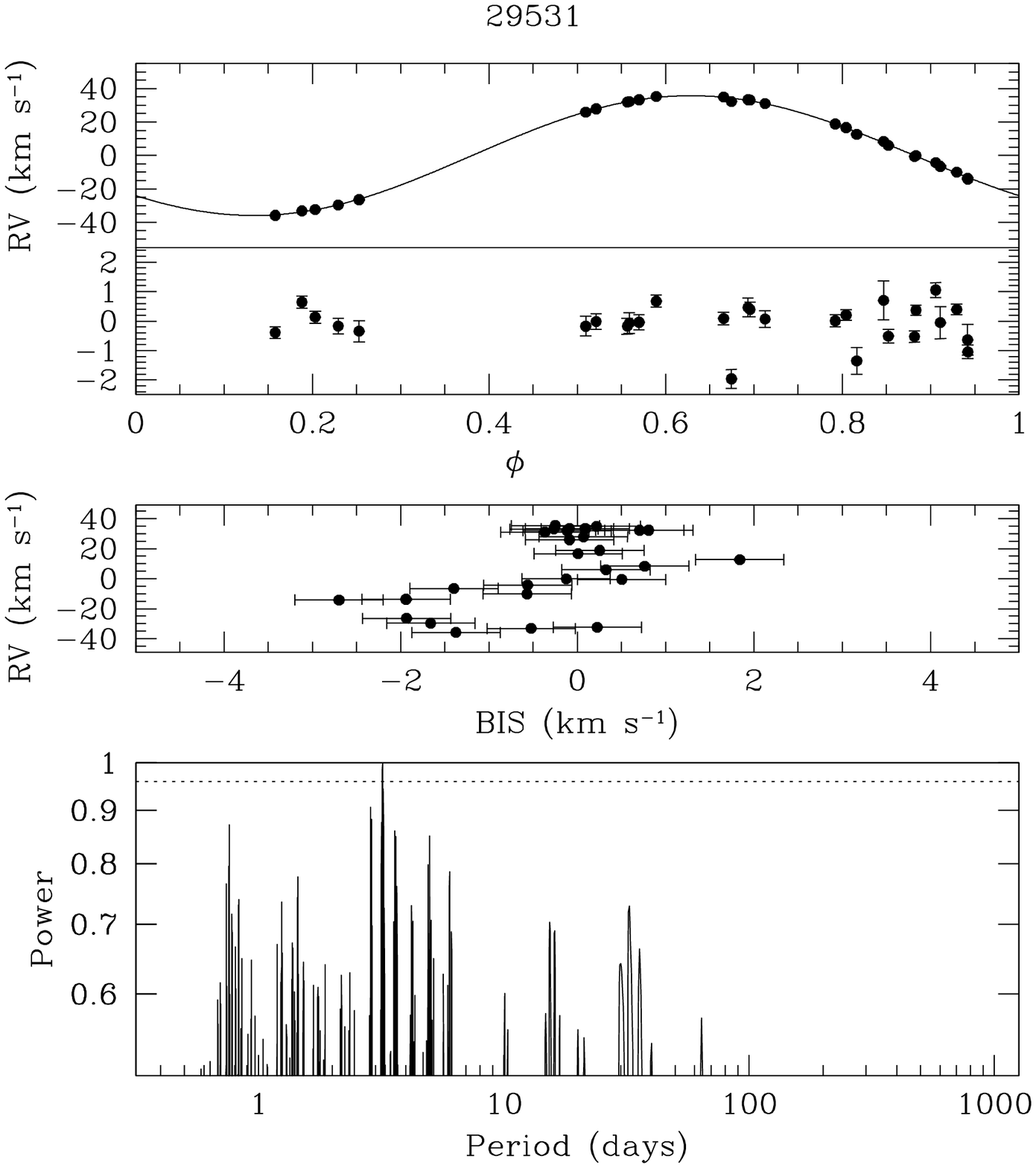}
\caption{
Folded  RV  curve with best fit  model  and 
residuals  (top), bisector
(middel) and periodogram (bottom) for star 29531.
The dashed horizontal line in the periodogram
panel denotes the GLS algorithm detection treshold
(p$\rm_{GLS,FAP}$) as defined in the text.
\label{fig:v29531}
}
\end{figure*}

\begin{figure*}
\centering
\includegraphics[width=16cm]{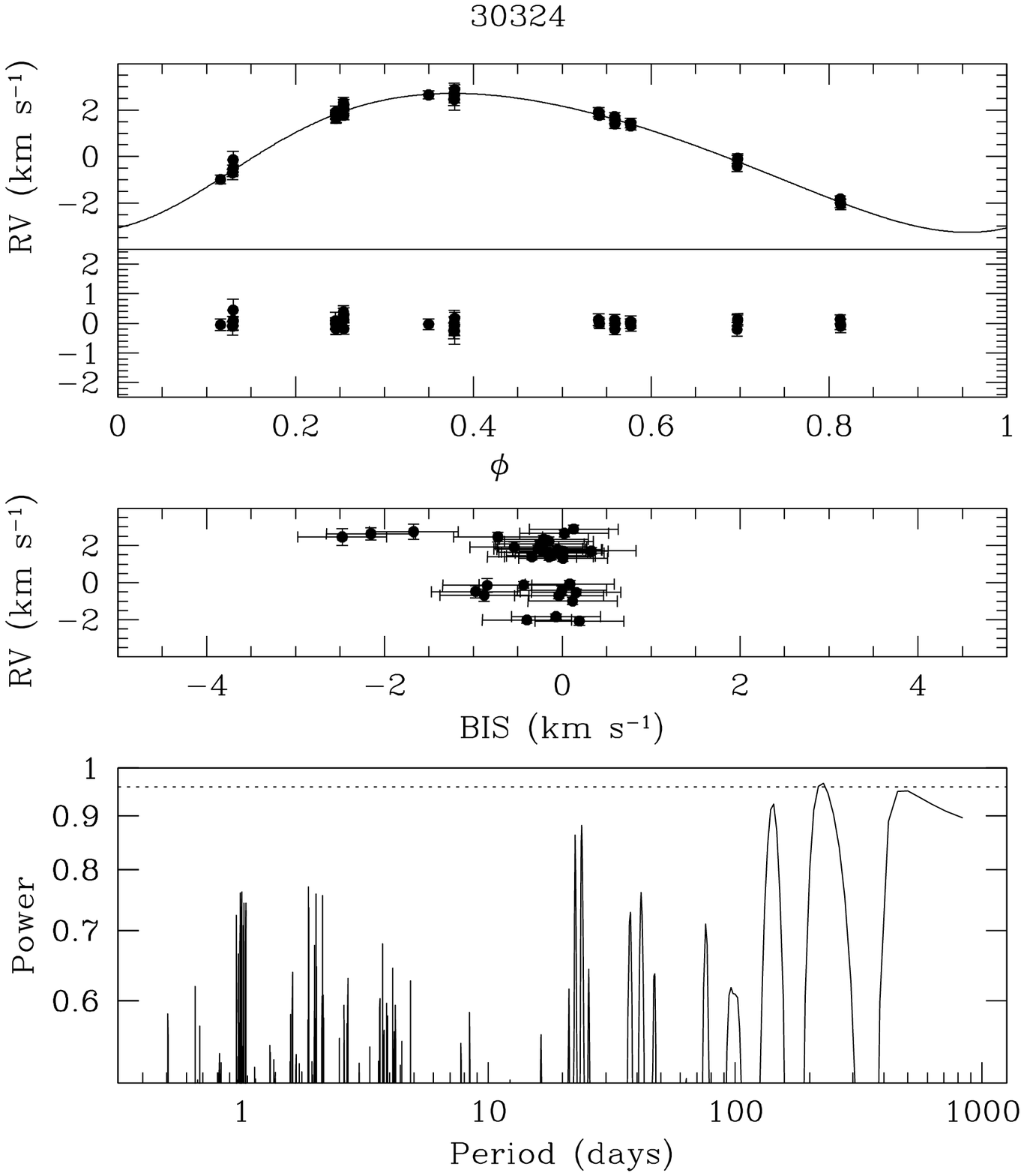}
\caption{
Folded  RV  curve with best fit  model  and 
residuals  (top), bisector
(middel) and periodogram (bottom) for star 30324.
The dashed horizontal line in the periodogram
panel denotes the GLS algorithm detection treshold
(p$\rm_{GLS,FAP}$) as defined in the text.
\label{fig:v30324}
}
\end{figure*}

\begin{figure*}
\centering
\includegraphics[width=16cm]{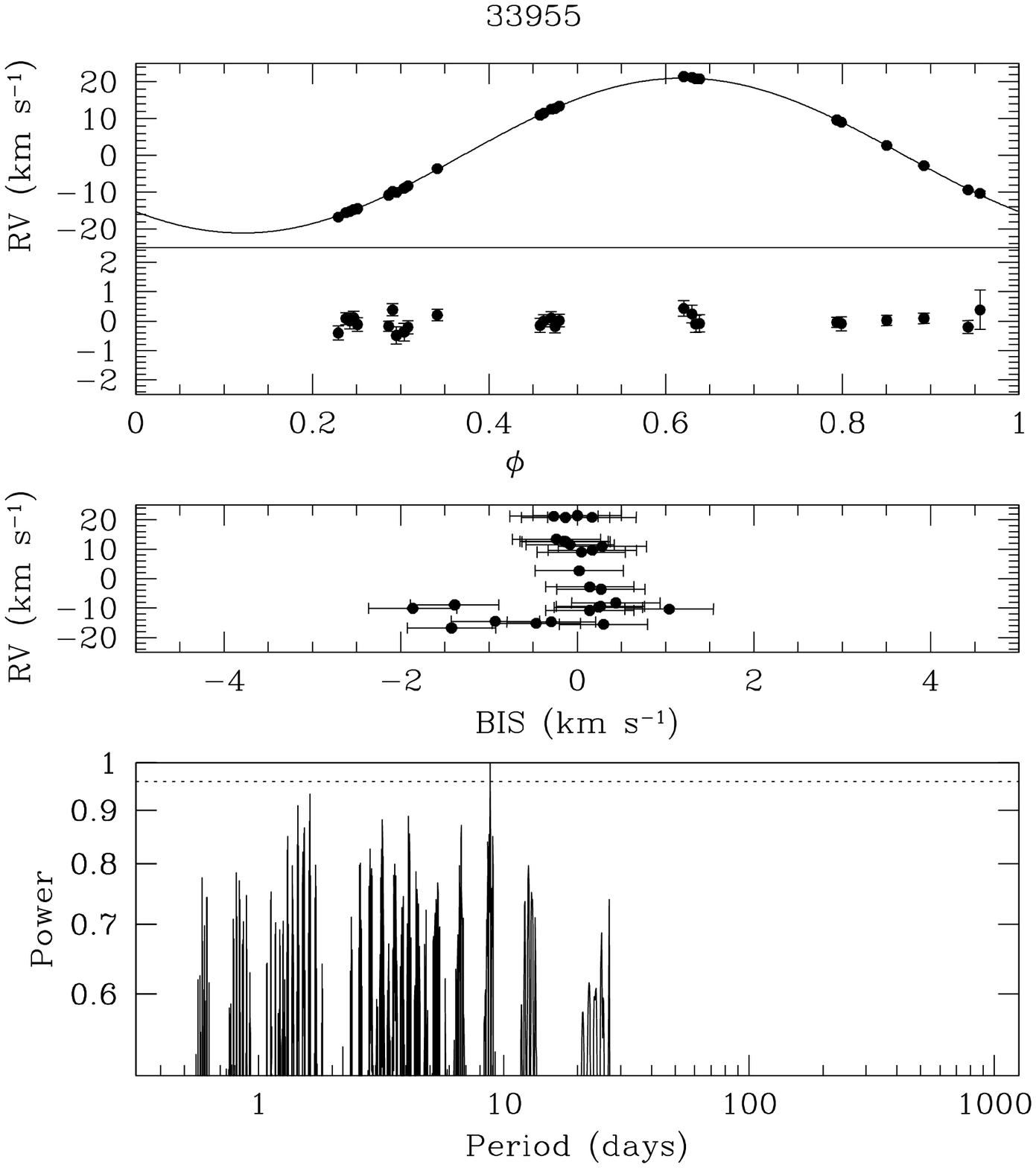}
\caption{
Folded  RV  curve with best fit  model  and 
residuals  (top), bisector
(middel) and periodogram (bottom) for star 33955.
The dashed horizontal line in the periodogram
panel denotes the GLS algorithm detection treshold
(p$\rm_{GLS,FAP}$) as defined in the text.
\label{fig:v33955}
}
\end{figure*}

\begin{figure*}
\includegraphics[width=16cm]{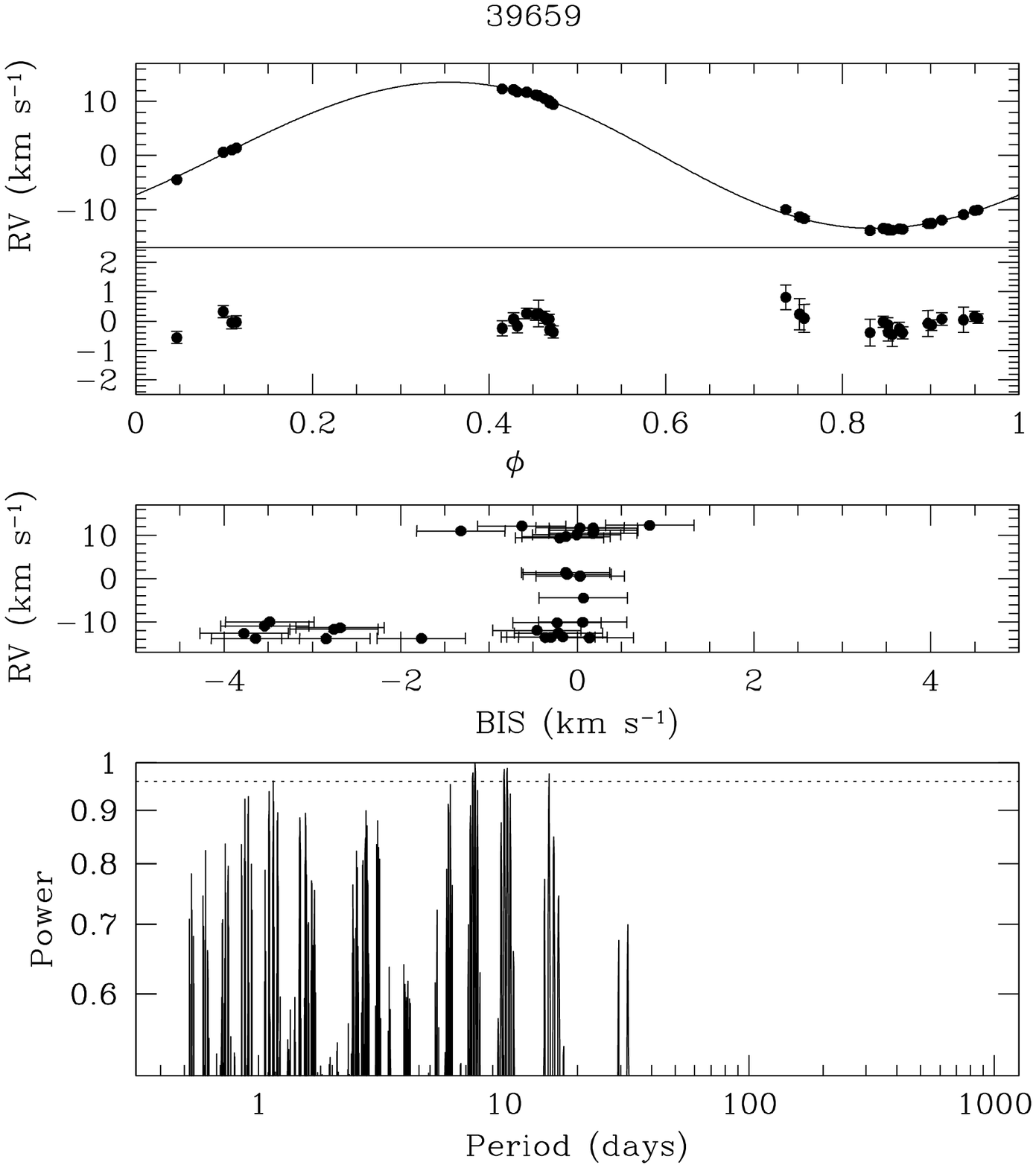}
\caption{
Folded  RV  curve with best fit  model  and 
residuals  (top), bisector
(middel) and periodogram (bottom) for star 39659.
The dashed horizontal line in the periodogram
panel denotes the GLS algorithm detection treshold
(p$\rm_{GLS,FAP}$) as defined in the text.
\label{fig:v39659}
}
\end{figure*}

\begin{figure*}
\includegraphics[width=16cm]{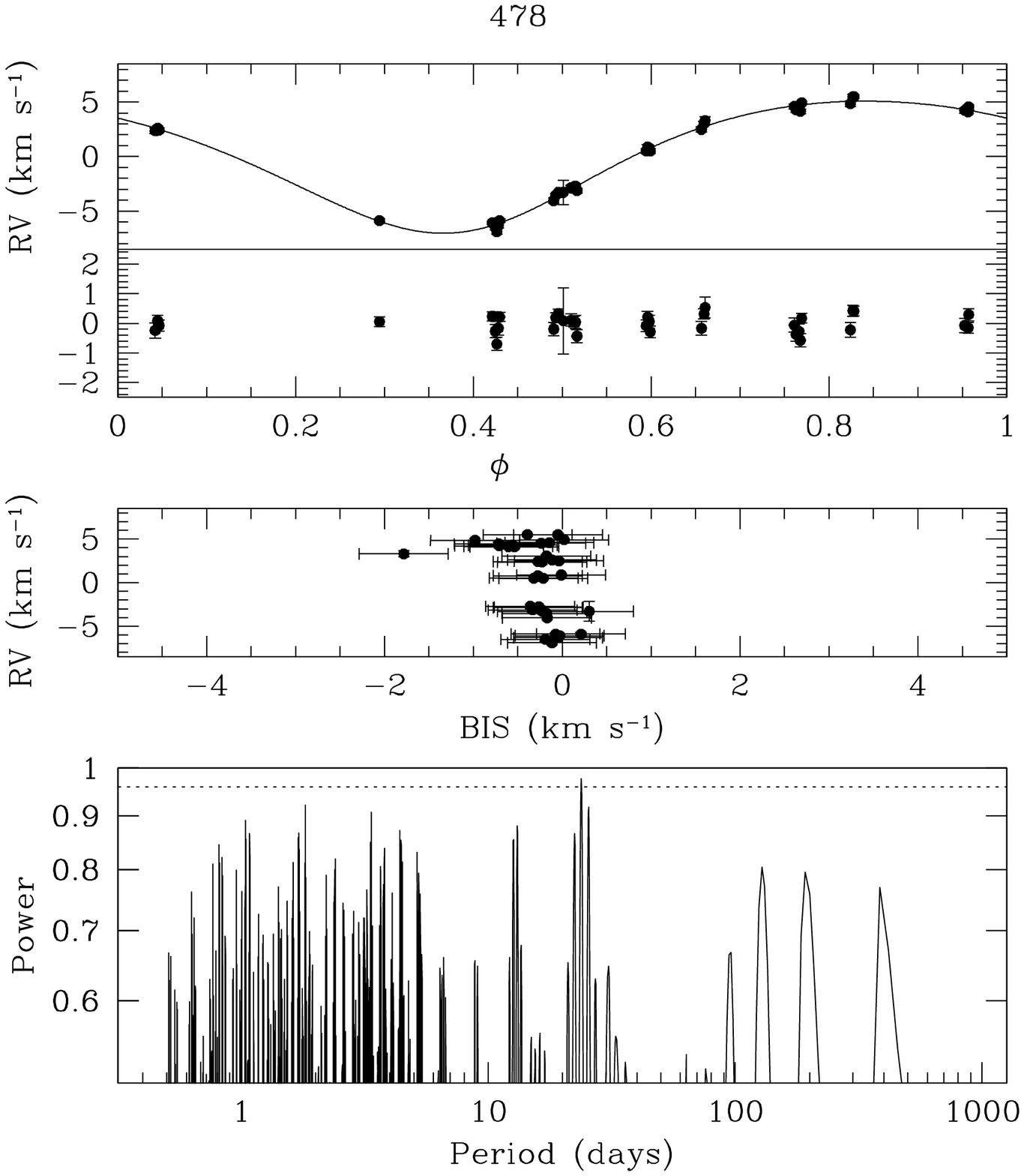}
\flushleft
\caption{
Folded  RV  curve with best fit  model  and 
residuals  (top), bisector
(middel) and periodogram (bottom) for star 478.
The dashed horizontal line in the periodogram
panel denotes the GLS algorithm detection treshold
(p$\rm_{GLS,FAP}$) as defined in the text.
\label{fig:v478}
}
\end{figure*}

\noindent
The radial velocity measurements  after subtraction of the barycentric
velocities, along  with the best  fit model, bisector diagram  and the
periodogram                 are                shown                in
Fig.~\ref{fig:v21659}-Fig.~\ref{fig:v478}.       Our      best-fit
parameters  are given  in  Table~\ref{tab:spectroscopic_binaries}. The
best-fit  models correspond to  values of  the reduced  $\chi$ squared
between $\sqrt{\chi_{r}^2}=0.8-2.5$. 

\noindent
The bisector  error was  calculated from  the dispersion  of the
distribution  of  bisector values  of  all  the  spectra, and  assumed
identical  for all  stars.  We  also checked  for  linear correlations
between the bisector  and radial velocity measurements calculating
the Pearson correlation coefficient.  For the six cases considered the
coefficient is comprised in between  -0.47 and 0.56 which denoted that
no strong correlations are found.  

\noindent
For those variables  that were considered cluster members  (4 out of 6
as shown in Table~\ref{tab:spectroscopic_binaries}) we also calculated
the  mass function and,  by assuming  for the  primary the  mass ($\rm
M_{*}$) obtained  by isochrone fit\footnote{We considered  the 3.5 Gyr
  isochrone   that   was   discussed    in   M09.}    we   report   in
Table~\ref{tab:spectroscopic_binaries}  a lower  limit on  the minimum
mass of the companion which is given by the following equation

\begin{equation}
\rm m\,sin\,i>\Big(M^2_{*}\frac{\tilde{K}^3\,P}{2\,\pi\,G}\Big)^{\frac{1}{3}}
\end{equation}

\noindent
the  value given by  the expression  on the  right represents  a lower
limit on  the minimum mass because  we neglected the term  on the mass
ratio $\rm  q=m/M_{*}$, which contributes as $\rm  (1+q)^{2/3}$ to the
numerator of  the term on  the right.  The  values we obtained  are in
between $\sim$90  $\rm M_{J}$ and $\sim$460 $\rm  M_{J}$.  The periods
instead are in between $\sim$3  days and $\sim$220 days.  We note that
the  attribution of  the  exact  period is  still  ambiguous for  some
objects. Among  this sample of  objects star 478  seems to have  a non
negligible eccentricity e=(0.16$\pm$0.03).  For star 29531 instead the
$\chi_r$ of  the fit is  2.5 and this  may suggest the presence  of an
additional companion.

\section{Comparison with previous surveys}
\label{sec:comparison}

So far four main surveys have been performed toward NGC~6253 to search
for variables. As reported in the introduciton De Marchi et al.~(2010)
and    Kaluzny    et     al.~(2014)    searched    for     photometric
variables. Anthony-Twarog  et al.~(2010) and   M11 searched for radial
velocity variables.

\noindent
We observed the  detached eclipsing binary V23 (39883  in our catalog)
reported in  Kaluzny et  al.~(2014).  This object  is a  proper motion
member of NGC~6253 and sits close  to the turn-off. We obtained only 1
measurement   on  it   which   gave  a   radial   velocity  equal   to
(-52.96$\pm$0.18) km$\rm\,s^{-1}$.

\noindent
The star  16649 was classified in De  Marchi et al.  (2010)  as a long
period variable, likely cluster member  and located at around 7 arcmin
from the  cluster center. It  showed a photometric variability  at the
1$\%$ level.  This  object has been extensively observed  here.  A set
of 32 measurements have been  acquired.  It results that star 16649 is
also   a   radial   velocity    variable   and   it   is   listed   in
Table~\ref{tab:cluster_variables} as a  likely cluster member variable.
The mean radial velocity is -28.446 km$\rm\,s^{-1}$ and the RMS is 7.7
km$\rm\,s^{-1}$.  Membership  is estabilished  solely on the  basis of
radial  velocity.   The GLS  algorithm  flagged  out  this star  as  a
variable, but at the nominal period ($\sim$63 days) it presents a very
poor  phase coverage  and  no orbital  solution  was possible.   These
observations support the idea that this object is a long period binary
system.  It  is curious  to note  that in fact,  this object  was also
observed  by  Anthony-Twarog et  al.~(2010)  during  the Hydra  radial
velocity   survey  on   NGC6253.  Also   in  that   case   star  16649
(corresponding to  star 7592 in  their numeration) was found  a radial
velocity  variable star.  The  authors report  a mean  radial velocity
equal to -27.38 km$\rm\,s^{-1}$ and an RMS equal to 10 km$\rm\,s^{-1}$
measured out of a set of 3 measurements.

\noindent
The interesting star 55053 was  also listed in De Marchi et al.~(2010)
as a  long period  variable, likely cluster  member. It is  located at
around 5.2 arcmin from the cluster  center. In our study we obtained a
set  of 15  radial velocity  measurements on  it.  Interestingly, this
object  is  not a  radial  velocity variable,  at  least  down to  0.4
km$\,\rm\,s^{-1}$, the value  we calculated for the RMS.  In our study
it  is   classified  as  a   radial  velocity  non   variable  cluster
member. Membership is estabilished on the basis of radial velocity
given that the average velocity is -28.613 km$\rm\,s^{-1}$.

\noindent
No  matches  were  found  with  our previous  radial  velocity  survey
presented  in Montalto  et  al.~(2011), because  there  we focused  on
brighter objects.   

\noindent
The overlap with the  Anthony-Twarog et al.~(2010)
survey  is instead  substantial. We  counted a  total of  32  stars in
common among the two surveys which gives us the opportunity to further
check  for  long-term variability  for  these  objects. Comparing  our
radial velocities with the ones of  the authors we found in general an
excellent agreement.   

\noindent
Star  21659  is  an   interesting  variable  that  was  discovered  by
Anthony-Twarog et al.~(2010). The authors  report an RMS equal to 19.5
km$\,s^{-1}$ for this object (their  star 7495) and cosidered it among
their list  of five likely cluster  variable stars. In  our study this
object was  classified as well as  a cluster variable.  In addition to
that it is among the list  of objects for which we derived the orbital
solution, as  reported in Table~\ref{tab:spectroscopic_binaries}.  The
radial  velocity  semi-amplitude   we  found  (41.8  km$\rm\,s^{-1}$)  is
consistent with the RMS reported by Anthony-Twarog et al.~(2010).

\noindent
Star 8247  (7303 in  Anthony-Twarog et al.~2010)  was considered  as a
likely member cluster variable by  the authors which gave an RMS equal
to  16.4  km$\rm\,s^{-1}$.   In  our  data this  object  is  instead  not
variable, but we  note that our mean radial  velocity differs from the
one of  the authors  by 13 km$\rm\,s^{-1}$.   This object is  likely a
long period cluster variable.

\noindent
On the contrary  star 7470 in Anthony-Twarog et  al.~(2010, which does
not have an  entry in our own  catalog), is in the list  of our likely
cluster variables in  Table~\ref{tab:cluster_variables}, having an RMS
equal to  4.519 km$\rm\,s^{-1}$. Anthony-Twarog et  al.~(2010) did not
consider this  object as a  variable, but their  RMS is similar  to us
(3.24  km$\rm\,s^{-1}$),  quite   larger  than  their  average  level,
although probably at the limit of detectability.

\section{Discussion}
\label{sec:simulations}

\noindent
The survey  we presented in this  work is one of  the deepest searches
for binaries in an old open cluster ever performed so far. Milliman et
al.~(2014),  based  on  their   on  going  long-term  radial  velocity
monitoring  of  the  old  open  cluster NGC~6819  concluded  that  the
fraction of binaries with periods  less than 10$^{4}$ days is equal to
22$\,\%\,\pm\,$3$\%$.  Our total frequency is lower as reported below,
but it is  also based on a 2 year survey  while Milliman et al.~(2014)
are monitoring NGC~6819 since 17  years. Binaries surveys in the field
(albeit  including a variety  of detection  methods) suggest  that the
total  multiplicity  fraction  around  field stars  is  around  54$\%$
(e. g. Raghavan et al.~2010).

\begin{figure*}
\centering
\includegraphics[width=16cm]{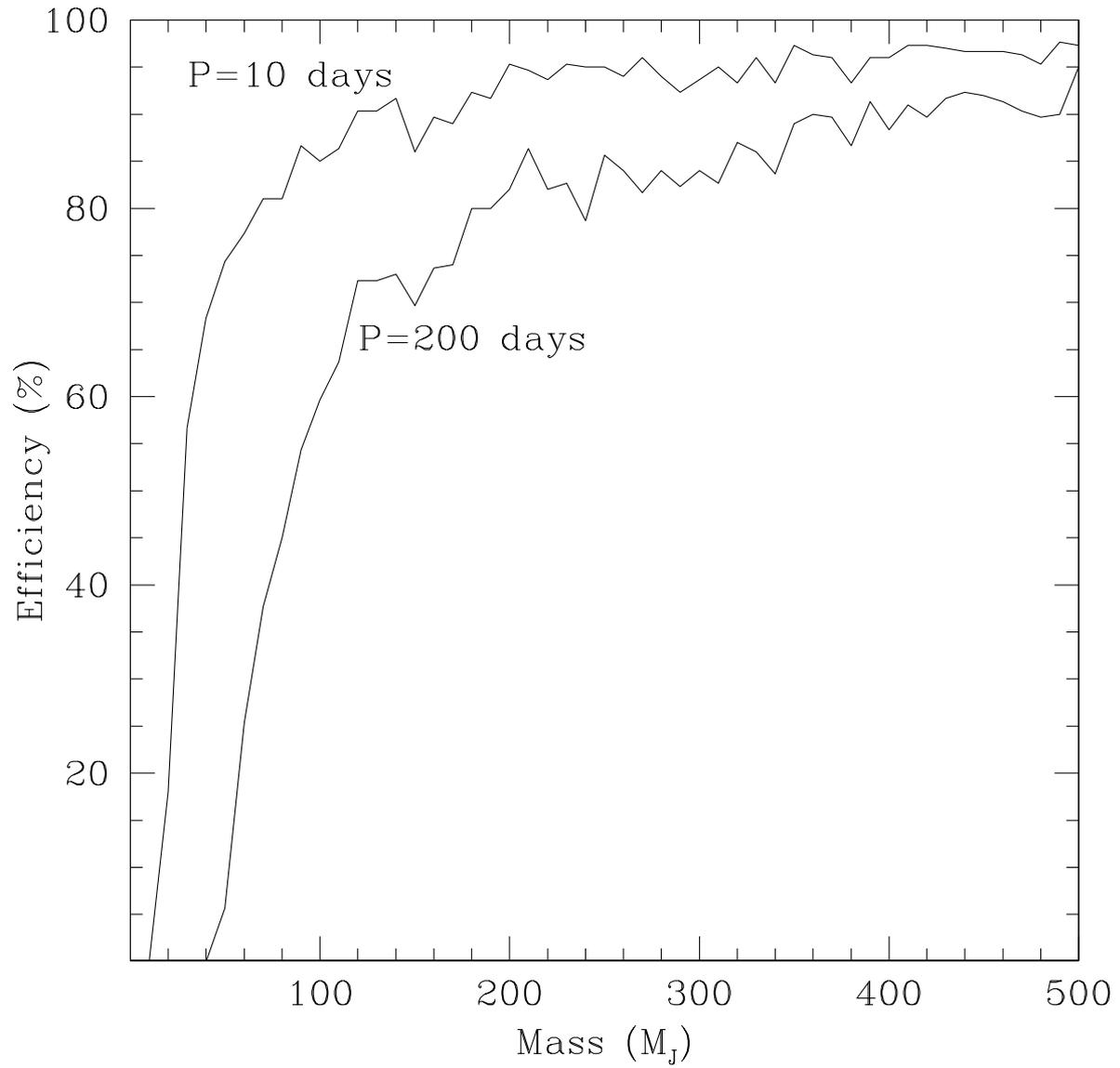}
\caption{
Survey detection tresholds as a function of the companion mass and period.
\label{fig:efficiency}
}
\end{figure*}

To understand our detection limits and binary fractions we performed a
set    of   simulations.     Similarly   to    what   was    done   in
Sect.~\ref{sec:search}, we  created a mock sample  of measurements for
each    observed   star,   assuming    the   baseline    noise   level
$\rm\sigma_{ref}$ and injecting an  artificial signal into the data to
mimick the presence of a  companion star. We randomized both the phase
and the inclination of the orbits  and considered 25 bins in true mass
(between 20 and 500 $\rm M_{J}$) and an orbital period equal either to
10 days  or to 200  days. We therefore applied  the 5$\rm\sigma_{ref}$
treshold  calculating  the  detection  efficiency in  each  mass  bin.
Figure~\ref{fig:efficiency}, shows the detection efficiency curves for
the two periods considered as a function of the companion mass.  Those
curves are  essentially average efficiencies over the  whole sample of
stars monitored. The simulations indicate that at a period of 10 days,
objects  with  masses down  to  $\sim$30  M$\rm_{J}$  could have  been
detected with  a $\sim$50$\%$ efficiency.  At  90$\%$ confidence limit
we  expect   to  be   able  to  detect   stars  with  at   least  0.15
M$\rm_{\odot}$.   
Considering  a  period of  200  days, a  50$\%$
detection  efficiency corresponds to  $\sim$0.1 M$\rm_{\odot}$,  and a
90$\%$ detection efficiency to $\sim$0.4 M$\rm_{\odot}$.  

Looking     at    the     results     we    obtained     in
Tab.~\ref{tab:spectroscopic_binaries},  we  conclude that  essentially
for  all the  detected  objects, the  detection  efficiency should  be
beyond  80$\%$. The  only exception  is star  30324 which  has  a long
period (223  days) and  a low  minimum mass (92  M$_{\rm J}$)  and may
indeed lie  in a  $\sim50\%$ efficiency region.   If the true  mass of
this object is close to the  limit indicated by its minimum mass, this
would indicate that a correcting factor  at least equal to 2 should be
applied  to  derive the  true  frequency  of  stars with  long  period
companions. This  may also be supported  by the fact that  for most of
the  variable objects we  detected it  was not  possible to  derive an
orbital solution.

\noindent
Considering  therefore only  the  sample of  close-in cluster  members
variables for  which an orbital  solution was found, we  conclude that
the frequency of binaries down  to the hydrogen burning limit and with
periods up to $\sim$20 days is around (1.5$\pm$1.3)$\%$, that is 3 out
200 members, in  the upper main sequence of  NGC~6253, while the total
binary  frequency is equal  to (13$\pm$3)$\%$.   The  errors were
obtained considering a binomial distribution.

\noindent
Comparatively, in  M11 we  obtained a frequency  of binaries  equal to
(29$\pm$9)$\%$.   These  binaries   were  all   flagged  out   with  a
5$\rm\sigma$ detection treshold like  in this work and were considered
likely short period binaries  given that the observations spanned only
a few  days.  The sample of  cluster stars (35) was  much smaller than
the one analyzed in this work which gives a larger uncertainty for the
estimated frequency. Even accounting for that, it appears however that
the frequency  among evolved stars  is larger (at a  2-$\sigma$ level)
than the total frequency of  binaries on the main sequence we obtained
above.

\section{Conclusions}
\label{sec:conclusions}

We described  an extensive radial  velocity survey toward the  old and
metal rich open cluster NGC~6253.  The survey was performed during two
seasons between April 2004 and July 2005 using the FLAMES spectrograph
monitoring a  total number  of 317  stars with a  median number  of 15
epochs per star.  A more limited follow up of  7 objects was performed
with the HARPS spectrograph in June 2011.

We obtained  a median  precision equal to  240 m$\rm\,s^{-1}$  for the
GIRAFFE   spectrograph,   180   m$\rm\,s^{-1}$   for   UVES   and   35
m$\rm\,s^{-1}$  for HARPS,  working in  a magnitude  range  in between
14.8$<V<$16.5.   Among  the sample  of  monitored  stars,  59 are  now
classified as radial velocity and proper motion cluster members and do
not show any evidence  of variability.  Field contaminations was equal
to 38$\%$ of the total sample.

In total  we found  45 variable stars,  among which 25  are considered
cluster members and 20 field objects.  For 6 spectroscopic binaries we
obtained the orbital solution, which implied minimum masses in between
$\sim$90  M$\rm_{J}$-$\sim$460  M$\rm_{J}$   and  orbital  periods  in
between $\sim$3 days and $\sim$220 days.

The frequency of binaries down to the hydrogen burning limit and up to
20 days  orbital period is  found equal to (1.5$\pm$1.3)$\%$.   in the
upper main sequence  of NGC~6253, while the total  binary frequency is
equal to (13$\pm$3)$\%$.

The precisions  achieved by the  HARPS spectrograph are  sufficient to
further extend the exploration of the binary frequency well within the
substellar domain, and  to analyze the high mass  end of the planetary
domain (down to 4-5  M$\rm_J$).  The upcoming ESPRESSO spectrograph at
VLT  will represent  an even  more  powerful tool  to study  planetary
frequencies   in  stellar   clusters.    A  precision   of  around   5
m$\rm\,s^{-1}$  may be  expected with  similar exposure  times  on our
targets.  This  will push  further  our  exploration  well within  the
planetary domain, down to around half a jupiter mass.

\section*{acknowledgements}
We acknowledge  the support from  Funda\c{c}\~ao para a Ci\^encia  e a
Tecnologia   (FCT,  Portugal)   in   the  form   of  grant   reference
PTDC/FIS-AST/1526/2014.  MM acknowledges  the support from FCT through
the grant and SFRH/BDP/71230/2010.  This wark is based on observations
made with  ESO Telescopes  at the La  Silla Paranal  Observatory under
programme  ID  073.C-0251,  075.C-0245   and  on  data  products  from
observations  made  with  ESO  Telescopes  at  the  La  Silla  Paranal
Observatory under  programme ID  087.C-0497. The anonymous  referee is
also aknowledged for the useful comments and suggestions which help us
to further improve this manuscript.











\begin{table*}
\caption{Observations}
\label{tab:observations}
\centering
\begin{tabular}{c c c c}
\hline
Observing time & Exposure time & Instrument & Dates \\
\hline
58hr   & 900s-3600s  & UVES/GIRAFFE & 2/4/2004-15/7/2004 \\
38.5hr & 2020s-2700s & UVES/GIRAFFE & 3/5/2005-1/7/2005  \\
13.5hr & 680s-3600s  & HARPS        & 28-30/05/2011      \\
\hline                                   
\end{tabular}
\end{table*}

\begin{table*}
\caption{Cluster variables$\rm^{a}$}
\label{tab:cluster_variables}
\centering
\begin{tabular}{ccccccccc}
\hline
ID & RA(J2000) & DEC(J2000) & {\it B-V} & {\it V} & N. obs. & $\rm\overline{RV}$ (km $\rm s^{-1}$) & RMS(km $\rm s^{-1}$) & P($\%$)$\rm^{b}$  \\
\hline           
  39667 &   254.732063568  &   -52.736025614   &    0.845  &    15.137   &     42  &   -29.087   &    4.030    &      88.  \\
  29531 &   254.699632781  &   -52.704366462   &    0.828  &    15.628   &     28  &   -27.300   &   24.509    &      89.  \\
  33955 &   254.700423295  &   -52.630001667   &    0.832  &    15.990   &     26  &   -29.945   &   13.486    &      90.  \\
  31617 &   254.689311198  &   -52.669417089   &    0.831  &    15.797   &     11  &   -28.108   &    1.725    &      91.  \\
  39854 &   254.781010467  &   -52.723246777   &    0.828  &    15.189   &      3  &   -30.762   &   10.677    &      94.  \\
  43081 &   254.767887749  &   -52.693853688   &    0.835  &    15.119   &     17  &   -30.744   &   11.849    &      93.  \\
  31834 &   254.734726224  &   -52.665662838   &    0.810  &    16.068   &     25  &   -26.430   &    3.467    &      82.  \\
\hline           
  30324 &   254.746004485   &  -52.690480237   &    0.814   &   15.940    &    41  &   -28.595   &    1.455     &     $-$  \\
  21659 &   254.876860986   &  -52.797707300   &    0.825   &   15.814    &    37  &   -29.166   &   30.154     &     $-$  \\
  39831 &   254.854747288   &  -52.724276741   &    0.789   &   15.464    &    36  &   -26.177   &   29.672     &     93.  \\
    $-$ &   254.871791667   &  -52.737444444   &    0.822   &   15.010    &    19  &   -29.012   &    2.035     &     $-$  \\
  35905 &   254.658158455   &  -52.596792279   &    0.866   &   16.213    &     6  &   -28.896   &    7.217     &     24.  \\
    $-$ &   254.979041667   &  -52.729833333   &    0.844   &   15.769    &     7  &   -28.434   &    4.519     &     $-$  \\
  24648 &   254.756000212   &  -52.791118012   &    0.913   &   16.428    &    39  &   -26.037   &    6.109     &      0.  \\
  49148 &   254.607856534   &  -52.777415622   &    0.846   &   15.550    &     7  &   -25.939   &    2.451     &     $-$  \\
  34667 &   254.708667366   &  -52.617760801   &    0.865   &   16.456    &    31  &   -29.062   &   13.748     &     85.  \\
  16649 &   254.892499238   &  -52.618200382   &    0.856   &   16.040    &    32  &   -28.446   &    7.678     &     $-$  \\
  30290 &   254.850999373   &  -52.690813347   &    0.825   &   15.544    &    41  &   -26.183   &   19.692     &     90.  \\
    $-$ &   254.794625000   &  -52.807416667   &    0.847   &   16.324    &    39  &   -29.067   &    3.014     &     $-$  \\
  14466 &   254.902801435   &  -52.714731958   &    0.854   &   14.774    &    28  &   -26.381   &    1.467     &     $-$  \\
  27965 &   254.847398993   &  -52.732799296   &    0.812   &   15.812    &    19  &   -28.740   &    5.294     &     70.  \\
  27341 &   254.863111261   &  -52.743533585   &    0.819   &   15.658    &    16  &   -26.352   &    9.431     &     85.  \\
  49667 &   254.600056580   &  -52.710400816   &    0.828   &   15.794    &     6  &   -30.035   &    6.405     &     $-$  \\
  41026 &   254.645602194   &  -52.641141858   &    0.811   &   14.869    &    22  &   -31.695   &    1.350     &      6.  \\
    $-$ &   254.977791667   &  -52.770055556   &    0.825   &   15.617    &     6  &   -28.714   &    2.009     &     $-$  \\
\hline
\end{tabular}
\newline
a -- $\,\,$Below the horizontal line membership is estabilished only on the basis of radial velocities.
\newline
b -- $\,\,$Proper motion membership probability.
\end{table*}

\begin{table*}
\caption{Field variables}
\label{tab:field_variables}
\centering
\begin{tabular}{cccccccccc}
\hline
ID & RA(J2000) & DEC(J2000) & {\it B-V} & {\it V} & N. obs. & $\rm\overline{RV}$ (km $\rm s^{-1}$) & RMS(km $\rm s^{-1}$) & P($\%$)\\
\hline
  31505  &   254.687118376  &   -52.671306636   &    0.864  &    15.728   &     41   &  -25.258   &    4.116    &      91. \\
    478  &   254.923022901  &   -52.809584139   &    0.879  &    14.783   &     35   &  -55.207   &    4.216    &     $-$  \\
  40696  &   254.778223506  &   -52.662991026   &    0.807  &    15.255   &     23   &  -24.845   &    5.866    &      95. \\
  28889  &   254.782822852   &  -52.715868624    &   0.824  &   15.701    &    29    &  -24.585   &    1.415    &      90. \\
  27108  &   254.786303150  &   -52.747918476   &    0.821  &    15.426   &     41   &  -23.120   &    3.147    &      89. \\
  28017  &   254.789168050  &   -52.731954093   &    0.814  &    15.436   &     25   &  -36.817   &   17.897    &      92. \\
   9156  &   254.897722920  &   -52.620682013   &    0.875  &    16.213   &      6   &   -2.398   &    1.442    &     $-$  \\
    $-$  &   254.956791667  &   -52.780388889   &    0.851  &    16.056   &     25   &  -24.514   &    7.795    &     $-$  \\
  39659  &   254.842084425  &   -52.736277043   &    0.768  &    15.266   &     30   &  -28.756   &   10.886    &      28. \\
  49446  &   254.621343512  &   -52.739416289   &    0.905  &    15.163   &     40   &  -12.317   &    4.569    &     $-$  \\
   9756  &   254.913463429  &   -52.608067153   &    0.910  &    16.263   &      8   &  -71.134   &   66.779    &     $-$  \\
  28733  &   254.790227260  &   -52.718374337   &    0.795  &    15.830   &     27   &  -26.188   &    7.492    &      56. \\
  23507  &   254.713571036  &   -52.811072103   &    0.808  &    15.779   &     36   &  -25.407   &    5.046    &     $-$  \\
  31512  &   254.810969168  &   -52.670869985   &    0.822  &    15.562   &     36   &  -32.037   &    1.814    &      90. \\
    $-$  &   254.933625000  &   -52.587555556   &    0.889  &    16.047   &      2   &  -21.799   &    2.256    &     $-$  \\ 
  10638  &   254.903502113  &   -52.587731891   &    0.946  &    15.490   &      8   &  -11.044   &    5.990    &     $-$  \\ 
    $-$  &   254.865708333  &   -52.674027778   &    0.885  &    16.024   &     37   &  -61.851   &    8.402    &     $-$  \\ 
  23848  &   254.736362325  &   -52.805171881   &    0.835  &    16.104   &     14   & -100.238   &    6.578    &     $-$  \\
  24548  &   254.742155375  &   -52.792685121   &    0.800  &    16.005   &     23   &  -93.670   &    1.698    &       0. \\
    $-$  &   254.769458333  &   -52.707944444   &    0.833  &    15.427   &      5   &  -23.596   &    1.624    &     $-$  \\
\hline
\end{tabular}
\end{table*}

\newpage

\begin{table*}
\caption{Spectroscopic binaries}
\label{tab:spectroscopic_binaries}
\centering
\begin{tabular}{cccccccccc}
\hline
ID & Period (days) & $\rm \tilde{K}$(m $\rm s^{-1}$) & e$\,$cos$\rm \,\omega$ &  e$\,$sin$\rm \,\omega$ &  $\rm t_{MAX}$ (HJD-2450000) & $\gamma$(km $\rm s^{-1}$) & P($\%$) & m$\,$sin$\,$i ($\rm M_{J}$) & $\chi_{red}$ \\
\hline
29531                  &   3.204288  &  35800$_{-79}^{+79}$   &  -0.0018$_{-0.0033 }^{+0.0033}$  &  -0.00788$_{-0.0027}^{+0.0019}$ &  3201.39496$_{-0.0028 }^{+0.0013}$ &  -27.30048$_{-0.047}^{+0.075}$  &   89 & 279 &  2.5 \\                
30324                  & 224.118409  &   3000$_{-146}^{+134}$ &  -0.091$_{-0.051 }^{+0.021 }$    &  -0.12344$_{-0.0277}^{+0.0099}$ &  3561.65496$_{-0.71}^{+0.83}$      & -28.59548$_{-0.074}^{+0.062}$  &   $-$ &  92 & 0.8 \\          
21659                  &   9.967883  &  41800$_{-70}^{+65}$   &  -0.0047$_{-0.0012 }^{+0.0010 }$ &   0.00345$_{-0.0025}^{+0.0015}$ &  3522.36771$_{-0.0024}^{+0.0017}$   &  -29.16622$_{-0.052}^{+0.027}$ &   $-$ & 466 & 1.3 \\
33955                  &   8.808062  &  21000$_{-65}^{+147}$  &  -0.0015$_{-0.0036 }^{+0.0054 }$ &  -0.00248$_{-0.0039}^{+0.0033}$ &  3552.49403$_{-0.0035}^{+0.0047}$   & -29.94483$_{-0.085}^{+0.043}$  &   90 & 220  & 1.0 \\         
39659$\rm^{a}$ &   7.651044  &  13500$_{-119}^{+171}$ &   0.0027$_{-0.017 }^{+0.0061 }$  &   0.02813$_{-0.0113}^{+0.0049}$ &  3144.08394$_{-0.018}^{+0.013}$    & -28.75597$_{-0.098}^{+0.058}$   &   28 & 146 & 1.2  \\     
  478$\rm^{a}$ &  23.846401  &   6100$_{-60}^{+81}$   &  -0.1612$_{-0.0093 }^{+0.010}$   &  -0.03923$_{-0.0080}^{+0.0094}$ & 3149.34776$_{-0.077}^{+0.021}$     & -55.20715$_{-0.053}^{+0.053}$  &    $-$ & 102 & 1.5 \\  
\hline
\end{tabular}
a -- $\,\,$Field star
\end{table*}

\clearpage

\newpage

\bsp	
\label{lastpage}
\end{document}